\documentclass[11pt]{article}

\usepackage{arxiv}
\usepackage{amssymb,amsmath,amsthm}

\usepackage{bm}
\usepackage[colorlinks,citecolor=blue]{hyperref}
\usepackage{subcaption}
\usepackage{multirow}
\usepackage{mathtools}
\usepackage{threeparttable}
\usepackage{booktabs}
\usepackage{blkarray, bigstrut} %
\usepackage{arydshln}
\usepackage{rotating}
\usepackage[numbers]{natbib}
\begin{document}
\title{Double robust estimation of partially adaptive treatment strategies}
\author{Denis Talbot \\
    denis.talbot@fmed.ulaval.ca \\
    D\'epartement de m\'edecine sociale et pr\'eventive \\	Universit\'e Laval 
\And
 Erica EM Moodie \\
 Department of Epidemiology and Biostatistics\\
 McGill University
 \And
 Caroline Diorio\\
 Axe oncologie, Centre de recherche du CHU de Qu\'ebec\\
 Universit\'e Laval
}

\maketitle

\begin{abstract}
Precision medicine aims to tailor treatment decisions according to patients’ characteristics. G-estimation and dynamic weighted ordinary least squares (dWOLS) are double robust statistical methods that can be used to identify optimal adaptive treatment strategies. They require both a model for the outcome and a model for the treatment and are consistent if at least one of these models is correctly specified. It is underappreciated that these methods additionally require modeling all existing treatment-confounder interactions to yield consistent estimators. Identifying partially adaptive treatment strategies that tailor treatments according to only a few covariates, ignoring some interactions, may be preferable in practice. It has been proposed to combine inverse probability weighting and G-estimation to address this issue, but we argue that the resulting estimator is not expected to be double robust. Building on G-estimation and dWOLS, we propose alternative estimators of partially adaptive strategies and demonstrate their double robustness. We investigate and compare the empirical performance of six estimators in a simulation study. As expected, estimators combining inverse probability weighting with either G-estimation or dWOLS are biased when the treatment model is incorrectly specified. The other estimators are unbiased if either the treatment or the outcome model are correctly specified and have similar standard errors. Using data maintained by the Centre des Maladies du Sein, the methods are illustrated to estimate a partially adaptive treatment strategy for tailoring hormonal therapy use in breast cancer patients according to their estrogen receptor status and body mass index. R software implementing our estimators is provided.
\end{abstract}

\quad

\begin{center}
\end{center}

\section{Introduction}
\label{s:intro}

Precision medicine, sometimes also called personalized medicine, is a medical approach that aims to improve patients' outcomes by tailoring treatment decisions taking into account their genes, environments and lifestyles (United States Food and Drug Administration, 2018). This approach has garnered increasing attention over recent years. Notably, multiple countries have put forward initiatives to accelerate research on precision medicine, including the United States, Canada and various European countries. Adaptive treatment strategies (ATSs), or dynamic treatment regimes, is a type of precision medicine. ATSs aim to determine time-dependent treatment decision rules that use prior patients' information to optimize a clinical outcome among patients sharing similar characteristics. 

Multiple statistical techniques have been proposed for estimating ATSs (for example, Watkins, 1989; Murphy, 2003; Robins, 2004; Wallace and Moodie, 2015). G-estimation (Robins, 2004) and dynamic  weighted ordinary least squares (dWOLS; Wallace and Moodie, 2015) have the advantage of offering some robustness to statistical modeling errors. These methods require the specification of two models: a treatment model that relates potential confounders to the treatment decisions, and an outcome model that relates potential confounders and treatments to the outcome. The outcome model is further divided in a treatment-free component that includes only potential confounders, and a ``blip'' component that features terms related to the treatment effect and possible effect modification according to tailoring variables. The blip is the statistical quantity of interest for determining optimal decision rules. G-estimation and dWOLS yield consistent estimators of the blip's parameters if either the treatment model or the treatment-free component is correctly specified, but not necessarily both, a property known as double robustness. The blip must, however, be correctly specified. Other double robust methods for estimating ATSs have been proposed (for example, Petersen et al, 2004; van der Laan and Petersen, 2007), but we focus on G-estimation and dWOLS because of their relative ease of implementation and interpretation within a regression-like framework.

G-estimation and dWOLS do not formally require including the same covariates in the blip component as in the treatment-free component. For dWOLS, any covariate interacting with the treatment in the blip must be included as main effect in the treatment-free component (that is, the ``hierarchy'' of the model must be preserved), while G-estimation has no specific requirement. In practice, it is common to include fewer covariates in the blip than in the treatment-free model. For example, when estimating an ATS for the treatment of type 2 diabetes, Simoneau et al (2020) included multiple potential confounders in the treatment-free component, but only glycemic control, body mass index, previous treatment and history of hypoglycemia in the blip component. Multiple reasons may motivate tailoring treatment decisions according to only a few variables. For instance, some variables may be confounders in the context of a given dataset, but are expected to be unavailable for treatment tailoring in the intended clinical context. Simplifying decision rules to facilitate their use in clinical practice or to improve statistical power are other examples of reasons to exclude some variables from the blip. 

A perhaps underappreciated possible consequence of excluding some potential confounders from the blip is that the blip may be incorrectly specified if some true treatment effect modification remains unmodeled. As mentioned previously, both G-estimation and dWOLS require the blip to be correctly specified for consistent estimation of its parameters. A solution to this problem has been proposed by van der Laan and Robins (2003). It combines G-estimation with inverse probability of treatment weighting (IPTW), where the IPTW is used to control the residual confounding bias attributable to the (voluntary) possible misspecification of the blip. However, we expect this solution is not fully double robust, since it relies on the correct specification of a treatment model to account for the blip misspecification.

In this paper, we propose alternative double robust estimators of partially adaptive treatment strategies (PATSs), where the treatment is tailored according to only some covariates, excluding potential true effect modifiers. In the next section, we introduce the notation and briefly review dWOLS estimation of an ATS. \textcolor{black}{The estimand of PATSs, its identification, and dWOLS estimators of this estimand are proposed in Section~\ref{s:PATS}}. Analogous G-estimation methods are available in Appendix~C. In Section~\ref{s:simulation}, we employ simulation studies to illustrate the potential bias of using standard ATSs estimators for estimating PATSs as well as the double robustness of our proposed estimators. In Section~\ref{s:application}, we estimate a PATS that aims to tailor hormonal therapy for treating breast cancer as a function of estrogen receptor status and body mass index (BMI) in data maintained by the Centre des Maladies du Sein. We conclude in Section~\ref{s:discussion} with a discussion of the results and perspectives for future research.  

\section{Notation and review of adaptive treatment strategy estimation} \label{s:notation}

\subsection{Notation}

Let $A_j$ be the exposure at time $j$ ($j = 1, ..., K$), $X_j$ the pre-treatment covariates at time $j$, and $Y$ the final outcome. Without loss of generality, we assume that a greater value of $Y$ represents a better clinical outcome and that $A_j = 0$ represents a reference value for the treatment (for example, a placebo or standard care). We use over- and underbars to represent the set of past and future values of a variable, respectively, both including the present. For example, $\bar{A}_j = \{A_1, ..., A_j\}$ and $\underline{A}_j = \{A_{j}, A_{j+1}, ..., A_K\}$. We further denote by $H_j$ the observed history before treatment $j$, including previous treatments; hence, $H_j = \{\bar{A}_{j-1}, \bar{X}_j\}$. Let $d_j$ denote a treatment decision rule, which is a function of the observed history. \textcolor{black}{For example, a decision rule at time $j$ ($d_j$) could be to give hormonal therapy ($A_j=1$) only to patients who have a normal BMI at time $j$ and whose breast cancer is either estrogen or progesterone receptor positive.} The goal of an ATS is to determine the optimal treatment decision rule $d_j^{opt}$ at each time $j$ in order to maximize the final outcome $Y$ \textcolor{black}{(e.g., log survival time)}. 
To define the causal parameter of interest more formally, we make use of the counterfactual framework. For example, $Y^{\bar{d}_K^{opt}}$ is the outcome that would have been observed if, possibly contrary to the fact, the treatment strategy had been optimal throughout the entire follow-up. Using this notation, the causal parameter of interest in an ATS is
\begin{align*}
\gamma_j(a_j, h_j) = \mathbb{E}\left[Y^{\bar{a}_{j-1}, a_j, \underline{d}_{j+1}^{opt}} - Y^{\bar{a}_{j-1},0,\underline{d}_{j+1}^{opt}}|H_j = h_j\right] = \mathbb{E}\left[Y^{\bar{a}_j,\underline{d}_{j+1}^{opt}} - Y^{\bar{a}_{j-1},0,\underline{d}_{j+1}^{opt}}|H_j = h_j\right].
\end{align*}
\noindent This is the blip to 0 function. It represents the effect of treatment $A_j = a_j$ compared to $A_j = 0$ for a given history $h_j$, assuming that the future treatment strategy is optimal ($\underline{a}_{j+1} = \underline{d}_{j+1}^{opt}$). \textcolor{black}{The optimal treatment strategy $d_j^{opt}$ can be formally defined recursively from $j = K, ..., 1$ as
\begin{align*}
d_j^{opt} = \underset{a_j}{argmax} \ \gamma_j(a_j, h_j).
\end{align*}
Continuing our earlier example, the blip $\gamma_j(a_j = 1, h_j)$ would represent the difference in the average log survival time $Y$, among subjects with a given history $h_j$, had they all been given hormonal therapy at time $j$ and the average log survival time had none of them been given hormonal therapy at time $j$, assuming in both situations that optimal treatment decisions would be taken at future time-points ($\underline{a}_{j+1} = \underline{d}_{j+1}^{opt}$). The optimal decision rule ($d_j^{opt}$) would then be to give hormonal therapy to a given individual if only if giving it increases their expected log survival time, that is, if $\gamma_j(a_j = 1, h_j) > 0$}. Note that the optimal future treatment strategy may not be the same under treatment history $\bar{a}_j$ as under treatment history $(\bar{a}_{j-1}, 0)$. In other words, $\underline{d}_{j+1}^{opt}$ may not represent the same treatment in $(\bar{a}_j,\underline{d}_{j+1}^{opt})$ as in $(\bar{a}_{j-1},0,\underline{d}_{j+1}^{opt})$, however in both cases the same treatment rule (or function) is applied; only the arguments to that function differ. 

Non-parametric identification of $\gamma_j(a_j, h_j)$ is possible from the observed data under the following assumptions:
\begin{itemize}
\item[(A.1)] \textit{Consistency}: If $\bar{A} = \bar{a}$ then $Y = Y^{\bar{a}}$. This assumption entails that the outcome of a given subject is not affected by the treatment of other subjects (absence of interference) and that there are not multiple versions of each treatment level $a$.
\item[(A.2)] \textit{Sequential exchangeability} (no unmeasured confounders): $(Y^{\bar{a}}, \underline{X}^{\bar{a}}_{j+1}) \coprod A_j | \bar{A}_{j-1}, \bar{X}_j$ for all $\bar{a}$ and $j = 1, ..., K$.
\item[(A.3)] \textit{Positivity}: $P(A_j = a_j|h_j) > 0$ for all $j = 1,..., K$, all $a_j$, and all $h_j$ such that $f(h_j) > 0$.   
\end{itemize}
\noindent If the positivity assumption fails to hold, it is still possible to estimate an optimal ATS among \textit{feasible} strategies, that is, among strategies that have been observed in the data. In practice, a parametric model is assumed for the blip function, such that we can write $\gamma_j(a_j, h_j; \psi_j)$. 

\subsection{Estimation of adaptive treatment strategies}

We now briefly present the dWOLS algorithm for estimating the blip $\gamma_j(a_j, h_j; \psi_j)$. More details can be found elsewhere (for example, see Wallace and Moodie, 2015) and a description of the G-estimation algorithm can be found in Appendix A, as well as in Robins (2004) or Moodie, Richardson and Stephens (2007). 


Estimation of $\psi_j$ using dWOLS is performed recursively. First, balancing weights such that $A_j \coprod H_j$ in the weighted data must be specified. Constructing such weights generally involves specifying a model for $\mathbb{E}[A_j|H_j]$; \textcolor{black}{in a binary treatment setting, this is typically accomplished by using a logistic regression}. Wallace and Moodie (2015) give a few examples of such balancing weights in the case of a binary treatment and observed that $w(A_j, H_j) = |A_j - \hat{\mathbb{E}}[A_j|H_j]|$ performs particularly well. These weights correspond to the overlap weights, which have been shown to yield a minimal variance weighted estimator of a certain average treatment effect under some assumptions (Li, Morgan and Zaslavsky, 2018; Li and Li, 2019). Schulz and Moodie (2020) propose various balancing weights for the multilevel and continuous treatment cases. Define the pseudo-outcome $\tilde{Y}_{iK} = Y_i$ and $\tilde{Y}_{ij} = Y_i + \sum_{k = j+1}^K [\gamma_k(d_k^{opt}, h_{ik}; \hat{\psi}_k) - \gamma_k(a_{ik}, h_{ik}; \hat{\psi}_k)]$ if $j \neq K$. Finally, a model for $\mathbb{E}[\tilde{Y}_j|A_j, H_j] = f_j(h_j; \beta_j) + \gamma_j(a_j, h_j; \psi_j)$ is specified, where $f_j(h_j; \beta_j)$ is a function relating $H_j$ to the outcome \textcolor{black}{known as the treatment-free component}, and its parameters are estimated using the weighted data. 

\textcolor{black}{In addition to the causal assumptions A.1-A.3, this dWOLS estimator requires i) the outcome to be modeled using a linear model, ii)~the correct specification of the blip components $\gamma_j(a_j, h_j; \psi_j)$, iii) the inclusion as main term in the treatment-free component $f_j(h_j; \beta_j)$ of any covariate included in the blip component $\gamma_j(a_j, h_j; \psi_j)$, and iv) the correct specification of either the treatment model $\mathbb{E}[A_j|H_j]$ or the treatment-free component $f_j(h_j; \beta_j)$. This fourth condition is the reason why dWOLS is a double robust estimator of the parameters of the blip (Wallace and Moodie, 2015). The intuition for the double robustness is as follows. If the treatment model $\mathbb{E}[A_j|H_j]$ is correctly specified, then the treatment at each time-point is independent of previous covariates and treatments in the weighted data, thus mimicking a randomized experiment with regard to the observed covariates. As a consequence, the treatment effect, which is encoded in the blip, can be estimated consistently. If the treatment model is misspecified, residual confounding can be present. However, adjustment for confounders in the outcome model $\mathbb{E}[\tilde{Y}_j|A_j, H_j]$ offers a second occasion to control bias. A more formal proof of the double robustness of dWOLS is provided in Appendix B of the Supplementary Material. } 


\section{Estimation of partially adaptive treatment strategies} \label{s:PATS}

We now turn our attention to the estimation of PATSs, that is, where the objective is to tailor treatment decisions only according to a subset of the measured history. To formally define the causal parameter of interest of a PATS, we partition the covariates history at each time-point in two disjoint subsets $H_j=(H_j^* \cup H_j^C)$, where $H_j^*$ are the covariates that are intended to be used for tailoring the treatment at time $j$ and $H_j^C$ are the other covariates. \textcolor{black}{This $^*-$notation is used in the following to refer to PATSs' quantities that are analoguous to those of the ATS. In our breast cancer example, we could define $H_j^*$ to be BMI at time $j$ and estrogen receptor status. All other pre-treatment variables would be in $H_j^C$, notably progesterone receptor status. In practice, the choice of the variables to include in $H_j^*$ would be based, for example, on their expected importance as tailoring variables and their availability in clinical practice.} The parameter of interest is defined as\textcolor{black}{
\begin{align}
\gamma_j^*(a_j, h_j^*) &= \mathbb{E}\left[Y^{\bar{a}_j,\underline{d}_{j+1}^{opt*}} - Y^{\bar{a}_{j-1},0,\underline{d}_{j+1}^{opt*}}|h_j^*\right], \label{eq.PATS0}
\end{align}
where $\underline{d}_{j+1}^{opt*}$ is the optimal treatment strategy from time $j+1$ onward when tailoring treatment only according to $H^*$. The optimal treatment at time $j$ when tailoring according to $H^*$, i.e. the optimal PATS, can be recursively defined from time $j = K, ..., 1$ as
\begin{align*}
d_j^{opt*} = \underset{a_j}{argmax} \ \gamma_j^*(a_j, h_j^*).
\end{align*}
As in the ATS setting, a parametric model would typically be assumed for $\gamma_j^*(a_j, h_j^*)$ and we can write $\gamma_j^*(a_j, h_j^*;\psi_j^*)$. So far, the quantities related to PATSs are very similary to those of ATSs. In fact, if $H_j^* = \varnothing$, then the PATS's estimand is the same as the ATS's estimand. However, neither the usual dWOLS or G-estimation estimators are generally consistent for the PATS's blip. Indeed, both estimators require the correct specification of the ATS's blip, but the PATS's blip can be volontarily misspecified because of the exclusion of true treatment-confounder interactions. 
} 

\textcolor{black}{Building on the dWOLS estimator of the ATS's blip $\gamma_j(a_j, h_j;\psi_j)$ presented in the previous section, we propose various novel estimators for the PATS's blip $\gamma_j^*(a_j, h_j^*;\psi_j^*)$. Similar G-estimators are presented in Appendix~C. First, the next subsection demonstrates the nonparametric identification of the PATS's blip. These results are central to the development of the estimators that are presented thereafter.}   

\subsection{Nonparametric identification}
\textcolor{black}{
The identification of PATSs we propose relies on the causal assumptions A.1-A.3. To demonstrate the nonparametric identification of PATSs, we first use the law of total expectations to rewrite the PATS's blip in Equation (\ref{eq.PATS0}) as 
\begin{align}
\gamma_j^*(a_j, h_j^*) &= \mathbb{E}_{H_j^C} \left\{ \mathbb{E}\left[Y^{\bar{a}_j,\underline{d}_{j+1}^{opt*}} - Y^{\bar{a}_{j-1},0,\underline{d}_{j+1}^{opt*}}|h_j\right] | h_j^* \right \}. \label{eq.PATS2}
\end{align}
\noindent Using this expression is essential because we are making an exchangeability assumption conditional on $H_j = \{\bar{A}_{j-1}, \bar{X}_j\}$, not on $H_j^*$. For the last time-point ($j = K$), this expression simplifies to $\mathbb{E}_{H_K^C} \left\{ \mathbb{E}\left[Y^{\bar{a}_K} - Y^{\bar{a}_{K-1},0}|h_K\right] | h_K^* \right\} = \mathbb{E}_{H_K^C}[\gamma_K(a_K, h_K)| h_K^*]$. As such, because $\gamma_K(a_K, h_K)$ is nonparametrically identified under our causal assumptions, so is $\gamma_K^*(a_K, h_K^*)$.}

\textcolor{black}{
The nonparametric identification for time-points $j < K$ is shown recursively. To simplify the presentation, we consider the case where $K = 2$. The causal effect of interest at the first time-point can thus be written as $\gamma_1^*(a_1, h_1^*) = \mathbb{E}_{H_1^C} \left\{\mathbb{E}\left[Y^{a_1,d_{2}^{opt*}} - Y^{0,d_{2}^{opt*}}|h_1\right] | H_1^* \right\}$. We have
\begin{align*}
\mathbb{E}[Y^{a_1, d_2^{opt*}}|h_1] &= \mathbb{E}[Y^{a_1, d_2^{opt*}}|a_1, h_1] \\
&= \mathbb{E}[Y^{a_1, d_2^{opt*}}|a_1, h_1] + \mathbb{E}[Y^{a_1, A_2}|a_1, h_1] + \mathbb{E}[Y^{a_1, 0}|a_1, h_1] - \mathbb{E}[Y^{a_1, A_2}|a_1, h_1] - \mathbb{E}[Y^{a_1, 0}|a_1, h_1] \\
&= \mathbb{E}[Y^{a_1, d_2^{opt*}}|a_1, h_1] + \mathbb{E}[Y|a_1, h_1] + \mathbb{E}[Y^{a_1, 0}|a_1, h_1] - \mathbb{E}[Y^{a_1, A_2}|a_1, h_1] - \mathbb{E}[Y^{a_1, 0}|a_1, h_1] \\
&= \mathbb{E}[Y|a_1, h_1] + \mathbb{E}\{\mathbb{E}[Y^{a_1, d_2^{opt*}} - Y^{a_1, 0}|H_2^*] -  \mathbb{E}[Y^{a_1, A_2} - Y^{a_1, 0}|H_2]|a_1, h_1\} \\
&= \mathbb{E}[Y + \gamma_2^*(d_2^{opt*}, H_2) - \gamma_2(A_2, H_2)|a_1, h_1].\\
\end{align*}
\noindent Because we have already shown that $\gamma_2^*(d_2^{opt*}, H_2)$ and $\gamma_2(A_2, H_2)$ are identified, we conclude that $\gamma_1^*(a_1, h_1^*)$ is also identified. Each of the estimators presented in the following sections can be seen as an empirical implementation of these identification formulas.}



\subsection{Combining dWOLS and Inverse probability of treatment weighting}

A first estimator of $\gamma_j^*(a_j, h_j^*;\psi_j^*)$ is based on the solution proposed by van der Laan and Robins (2003), Sections 6.5.2 and 6.5.3, in the more general context where the causal parameter of interest is expressed as a function of the density of the counterfactual outcomes. Their proposal consists of combining G-estimation with IPTW, where the IPTW is employed to account for the possible misspecification of the blip due to unmodeled true effect modification. We detail an analogous estimator combining IPTW with dWOLS, which we denote ``IPTW+dWOLS.'' Following Wallace and Moodie (2015), we show in Appendix~D that the solution to the IPTW+dWOLS estimator is also a solution the ``IPTW+G-estimation'' estimating equations in certain circumstances.  

\textcolor{black}{The algorithm for estimating the parameters of the PATS blip $\psi_j^*$ for $j = K, ..., 1$, is similar to the dWOLS algorithm used for estimating the parameters of the ATS blip $\psi_j$. Because the identification of $\gamma_j^*(a_j, h_j^*;\psi_j^*)$ involves $\gamma_j(a_j, h_j;\psi_j)$, the first step is to estimate $\psi_j$ with dWOLS.} Next, balancing weights $w^*(A_j, H_j^*)$ such that $A_j \coprod H_j^*$ in the weighted data must be specified. \textcolor{black}{The role of these weights is to control confounding due to the tailoring covariates $H_j^*$.} In addition, weights $\varpi(A_j, H_j, H_j^*) = \frac{P(A_j|H_j^*)}{P(A_j|H_j)}$ are computed. \textcolor{black}{This second set of weights further control for the confounding of non-tailoring covariates $H_j^C$. Again based on the identification formula, we build the following pseudo-outcomes: $\tilde{Y}_{iK}^* = Y_i$ and $\tilde{Y}_{ij}^* = Y_i + \sum_{k = j+1}^K [\gamma_k^*(d_k^{opt*}, h_{ik}^*; \hat{\psi}_k^*) - \gamma_k(a_{ik}, h_{ik}; \hat{\psi}_k)]$ if $j \neq K$}. Finally, specify a model for $\mathbb{E}[\tilde{Y}^*_j|A_j, H_j] = f_j(h_j; \beta_j) + \gamma_j^*(a_j, h_j^*; \psi_j^*)$ and estimate its parameters using the data weighted according to $w^*(A_j, H_j^*) \times \varpi(A_j, H_j, H_j^*)$. 

\textcolor{black}{This IPTW+dWOLS estimator of PATSs may give the impression of being double robust since confounding bias is controlled both through weighting and outcome-regression adjustment. However, because interactions between non-tailoring variables $H_j^C$ and treatment are ignored, the outcome model is expected to be misspecified. As a consequence, this approach relies on the correct specification of $\varpi(A_j, H_j, H_j^*)$ to eliminate the confounding bias due to the non-tailoring variables.} 

\subsection{Alternative estimators}

\textcolor{black}{We now propose two double robust alternative estimators of $\gamma_j^*(a_j; h_j^*; \psi_j^*)$. Both estimators are identical in all but their last step. These estimators proceed recursively for $j = K, ..., 1$. The first step is to estimate $\psi_j$ with dWOLS. The next step is to specify balancing weights $w(A_j, H_j)$ such that $A_j \coprod H_j$ in the weighted data. Note that these weights are the same as those used in the dWOLS estimator of ATSs. The same pseudo-outcomes $\tilde{Y}_{j}^*$ as in IPTW+dWOLS are then computed. A model for $\mathbb{E}[\tilde{Y}^*_j|A_j, H_j] = f(h_j; \beta_j) + \gamma_j^{\dagger}(a_j, h_j; \psi_j^{\dagger})$ is then fitted and its parameters are estimated using the weighted data. Under our causal assumptions, $\gamma_j^{\dagger}(a_j, h_j; \psi_j^{\dagger}) = \mathbb{E}\left[Y^{\bar{a}_j,\underline{d}_{j+1}^{opt*}} - Y^{\bar{a}_{j-1},0,\underline{d}_{j+1}^{opt*}}|h_j\right]$ which corresponds to the inner expectation of (\ref{eq.PATS2}). Note that $\gamma_j^{\dagger}(a_j, h_j; \psi_j^{\dagger})$ differs from $\gamma_j(a_j, h_j; \psi_j)$ because the former assumes that future treatment is tailored according to $\underline{H}_{j+1}^*$, whereas the latter assumes that future treatment is tailored according to $\underline{H}_{j+1}$. In addition, $\gamma_j^{\dagger}(a_j, h_j; \psi_j^{\dagger})$ differs from $\gamma_j^*(a_j, h_j^*; \psi_j^*)$ because they are not functions of the same variables ($h_j$ vs $h_j^*$).} 

The last remaining step is to compute the outer expectation of (\ref{eq.PATS2}).  A first approach for doing this is to estimate the density $f_{H_j^C|H_j^*}$ and integrate $\gamma_j^{\dagger}(a_j, h_j; \psi_j^{\dagger})$ over $H_j^C$. In simple situations, the nonparametric empirical estimator $\hat{f}_{H_j^C|H_j^*}(h_j^c|h_j^*) = \frac{\sum_{i=1}^n I(H_{ij}^C = h_j^c, H_{ij}^* = h_j^*)}{\sum_{i=1}^n I(H_{ij}^* = h_j^*)}$, where $I(\cdot)$ is the usual indicator function, could be used. Other density estimators would generally be required. We henceforth call this approach ``integrate  dWOLS'' and the analogous G-estimation approach as ``integrate G-estimation.''

An alternative implementation consists in computing the expectation through a linear model. More precisely, we calculate $Q(a, h_{ij}) = \gamma_j^{\dagger}(a, h_{ij}; \hat{\psi}^*)$ for all observations, that is, the predicted value of $\gamma_j^{\dagger}(A_j, h_{ij}; \psi_j^{\dagger})$ when setting $A_j = a$. Next, we regress the predicted $\gamma_j^{\dagger}(A_j, h_{ij}; \psi_j^{\dagger})$ on covariates $H_j^*$, thus fitting a model $\mathbb{E}[Q(a, H_j)|H_j^*] = \gamma^*(a, h_j^*; \psi^*_j)$. A clear advantage of this implementation is that it does not require estimating densities or computing integrals. \textcolor{black}{However, this approach requires the correct specification of the linear model that is used to compute the expectation.} This approach is called ``CE dWOLS'' and the analogous G-estimation approach is called ``CE G-estimation'' moving forward.

\subsection{Double robustness}

The intuition for the double robustness of ``integrate dWOLS'' and ``CE dWOLS'' is the same as that of dWOLS presented at the end of Section 2.2. More formally, the double robustness of our estimators can be demonstrated recursively. First, if considering the usual IPTW $w(A_j, H_j) = A_j/P(A_j = 1|H_j) + (1 - A_j)/P(A_j = 0|H_j)$, note that the double robustness of $\hat{\psi}^*_K$ and $\hat{\psi}_K$ (i.e. the treatment rule parameters from the final stage) is a direct consequence of the results of Kang and Schafer (2007) and Robins et al (2007) who demonstrate the double robustness of the ordinary least square estimator weighted according to IPTW, as long as an intercept term is included in the outcome (treatment-free) model. In Appendix~B, we provide a proof of the double robustness of $\hat{\psi}_K$ for the case where the weights $w(A_j, H_j) = |A_j - \mathbb{E}[A_j|H_j]|$ are used. The double robustness of $\hat{\psi}^*_K$ follows directly. To demonstrate the double robustness of $\hat{\psi}_j^*$ for $j < K$, assume that $\gamma^*_k(a_k, h_k^*; \psi_k^*)$ and $\gamma_k(a_k, h_k; \psi_k)$ for $k > j$ are consistently estimated. As a result, the mean of $\tilde{Y}^*_j$ for each $(a_j, h_j)$ is also consistent for $\mathbb{E}\left[Y^{\bar{a}_j,\underline{d}_{j+1}^{opt*}}|h_j\right]$. Appealing once more to the previous double robustness results, it follows that $\gamma_j^{\dagger}(a_j, h_j; \hat{\psi}_j^\dagger)$ is a double robust estimator for $\mathbb{E}\left[Y^{\bar{a}_j,\underline{d}_{j+1}^{opt*}} - Y^{\bar{a}_{j-1},0,\underline{d}_{j+1}^{opt*}}|h_j\right]$. Hence, $\hat{\psi}_j^*$ is also double robust for $\psi_j^*$. We conclude that consistent estimation of the parameters of $\gamma^*_j(a_j, h_j^*; \psi_j^*)$ at each time-point requires either the treatment model or the treatment-free model to be correctly specified, in addition to correct specification of the blips $\gamma^*_j(a_j, h_j^*; \psi_j^*)$, $\gamma_j(a_j, h_j; \psi_j)$ and $\gamma_j^\dagger(a_j, h_j; \psi_j^\dagger)$ with respect to included covariates (which may, by virtue of the PATS model, be a subset of all variables that interact with treatment), and the consistent estimation of the parameters of later blips. 

\subsection{Inferences}

Analytical calculation of confidence intervals for PATS is challenging because of the form of the parameter of interest. Alternatively, nonparametric bootstrap may be employed. However, this approach, as well as usual asymptotic variance estimators, would not be valid in cases where some $\gamma_j^*(a_j, h_j^*; \psi_j^*) = 0$ when $K > 1$ for $j \neq K$ and some $h_j^*$. Indeed, the optimal treatment strategy $\underline{d}_{j}^{opt*}$ is not uniquely defined in such situations, which yield estimators with non-regular limiting distribution (Robins, 2004). Even when $\gamma_j^*(a_j, h_j^*; \psi_j^*)$ is small relative to the sample size but nonzero, poor inferences may be produced by standard bootstrap because of the near non-regularity of the estimator (Moodie and Richardson, 2010). However, the $m$-out-of-$n$ bootstrap has been found to perform well in such settings (Chakraborty, Laber and Zhao, 2013; Simoneau et al, 2018). This type of bootstrap is similar to regular nonparametric bootstrap, except that $m < n$ observations are sampled at each replication. While the validity of this type of bootstrap relies on an appropriate choice of $m$, Chakraborty et al (2013) proposed a data-adaptive method for this choice that has been found to perform well. The data-adaptive choice determines $m$ based, in part, on an estimate of the extent to which the estimator is non-regular, which is determined from the data by the estimated proportion of subjects for whom there is no unique optimal choice of treatment (see Appendix F for details). We therefore recommend utilizing this data-adaptive $m$-out-of-$n$ bootstrap in general, although the regular nonparametric bootstrap may also be considered when nonregularity is not expected and the additional computational burden of the adaptive $m$-out-of-$n$ bootstrap is prohibitive. 

\section{Simulation study} \label{s:simulation}

\subsection{Scenarios}

\textcolor{black}{We conducted a simulation study whose main objectives were 1) to illustrate the double robustness of our proposed estimators (``integrate dWOLS'', ``CE dWOLS'', ``integrate G-estimation'' and ``CE G-estimation'') for estimating the parameters of PATSs, 2) to illustrate that ``IPTW+dWOLS'' and ``IPTW+G-estimation'' can produce biased results when the treatment model is misspecified and 3) to illustrate that ATS estimators can be biased for estimating the parameters of PATS. As a secondary objective, we also wanted to compare the empirical performance of the different PATS estimators. The current section focuses on scenarios with a single time-point.} Additional simulations in a two time-points setting are available in Appendix~E. The simulation scenarios are inspired by those presented in Wallace and Moodie (2015) and Wallace, Moodie and Stephens (2016).

Three different scenarios are considered. In all scenarios, $H = X = (X_1, X_2)$ are two pre-treatment covariates and both are effect modifiers. However, only $X_1$ is intended to be used to tailor treatment ($H^* = X_1$, $H^C = X_2$). In all three scenarios $X_1 \sim Bernoulli(p = 0.5)$ and $X_2 \sim Bernoulli(p = 0.5)$. In Scenarios 1 and 3, $A \sim Bernoulli(p = expit\{-0.5 + X_1 + 0.5X_2\})$; in Scenario 2, $A \sim Bernoulli(p = expit\{-0.5 + X_1 + 0.5X_2 + X_1 X_2\})$. In Scenario 1 and 2, $Y \sim N(0.25 X_1 + X_2 + A\{0.5 - X_1 + 1.5 X_2\}, 1)$; in Scenario 3, $Y \sim N(0.25 X_1 + X_2 + X_1 X_2 + A\{0.5 - X_1 + 1.5 X_2\}, 1)$. The true $\gamma_1^*(A, X_1)$ is $A(\psi_0^* + \psi_1^* X_1) = A(0.5 + 1.5\mathbb{E}[X_2|X_1] - X_1) = A(1.25 - X_1)$. The optimal PATS is thus $d^{opt*} = 1$, whether $X_1=0$ or $X_1 = 1$. Note that the optimal PATS varies according to covariates' values in the two time-points setting considered in Appendix~E.

The parameters $\psi_0^*$ and $\psi_1^*$ are estimated with ``IPTW+G-estimation'', ``IPTW+dWOLS'', ``integrate G-estimation'', ``integrate dWOLS'', ``CE G-estimation'', ``CE dWOLS'' in addition to usual implementations of G-estimation and dWOLS for ATS. For all estimators, the blip model includes only $X_1$, and the treatment and the treatment-free models only include main terms (no interactions). As such, in Scenario~1, both the treatment and the treatment-free models are correctly specified. In Scenario~2, only the treatment-free model is correctly specified. In Scenario~3, only the treatment model is correctly specified. 

A total of 1000 replications of each simulation scenario were generated for each sample sizes of $n = 100, 1000$ and $10~000$. We report below the estimated relative bias ((average estimate - true value)/true value $\times$ 100\%; Rel. bias) and standard deviation (SD) over the 1000 replications for each estimator in each combination of scenarios and sample size. The proportion of observations for which the optimal PATS was correctly identified and the expected loss (difference between the expected outcome under the true optimal PATS and the estimated optimal PATS) were also calculated within the same simulated data. In addition, we have explored the coverage of confidence intervals using the adaptive $m$-out-of-$n$ bootstrap proposed by Chakraborty et al (2013) in a two-time-points scenario with near non-regularity (see Appendix~F for details). Recall that non-regularity is not of concern in the single-stage setting, and so $m$-out-of-$n$ results are not presented for the simulations involving only a single treatment decision.

\subsection{Results}

The results of the main simulations are presented in Tables~1 and 2. Standard G-estimation and dWOLS, i.e. the approaches that use the incorrect (reduced) blip model with no adjustment for the partial nature of the ATS, produced estimates with non-negligible bias in all scenarios. On the other hand, all PATS estimators we introduced achieved unbiased estimation when the treatment model was correctly specified (Scenario~1 and 3). When the treatment model was misspecified, some bias remained for IPTW+G-estimation and IPTW+dWOLS, but not for the other PATS estimators. The standard deviation of the estimates were almost identical for all PATS estimators. The true optimal PATS was identified less often by standard G-estimation and dWOLS than the PATS estimators for $n = 100$ and $n = 1000$ in Scenarios~1 and 3, and for all sample sizes in Scenario~2. IPTW+dWOLS and IPTW+G-estimation also identified the optimal PATS less often than the other PATS estimators in Scenario~2 when $n = 100$ and $n = 1000$. For observations (``individuals'') whose true optimal PATS failed to be identified, the expected loss was always 0.25. Similar results were observed in the two time-point simulations, except that the difference in the ability to identify the optimal PATS between standard ATS estimators and the PATS estimators was more pronounced (see Appendix~E). The coverage of the confidence intervals produced using the adaptive $m$-out-of-$n$ bootstrap was approximately 95\% (between 95.2\% and 95.7\% depending on the parameter), as desired (Appendix~F).   

\begin{table}[h!] \centering 
  \caption{Bias and standard deviation of estimators of partially adaptive treatment strategies in three simulation scenarios} 
  \label{tab1}
	\resizebox{\textwidth}{!}{	
\begin{tabular}{@{\extracolsep{5pt}} lcccc|cccc|cccc} 
\\ \hline 
& \multicolumn{4}{c}{$n$ = 100} & \multicolumn{4}{c}{$n$ = 1000} & \multicolumn{4}{c}{$n$ = 10~000} \\
& \multicolumn{2}{c}{Rel. bias} & \multicolumn{2}{c}{SD} & \multicolumn{2}{c}{Rel. bias} & \multicolumn{2}{c}{SD}& \multicolumn{2}{c}{Rel. bias} & \multicolumn{2}{c}{SD}\\
Methods & $\psi_0$ & $\psi_1$ & $\psi_0$ & $\psi_1$ & $\psi_0$ & $\psi_1$ & $\psi_0$ & $\psi_1$ & $\psi_0$ & $\psi_1$ & $\psi_0$ & $\psi_1$ \\ 
\hline
\multicolumn{13}{c}{Scenario 1: correctly specified treatment and treatment-free models} \\
\hline
dWOLS           & $3.28$ & $10.7$  & $0.33$ & $0.49$ & $2.21$ & $9.29$ & $0.10$ & $0.14$& $1.88$ & $9.13$ & $0.03$ & $0.04$\\ 
G-est           & $3.22$ & $10.5$  & $0.33$ & $0.50$ & $2.19$ & $9.25$ & $0.10$ & $0.14$& $1.88$ & $9.13$ & $0.03$ & $0.04$\\ 
IPTW+dWOLS      & $1.30$ & $1.21$  & $0.32$ & $0.49$ & $0.41$ & $0.37$ & $0.10$ & $0.14$& $0.03$ & $0.11$ & $0.03$ & $0.04$\\ 
IPTW+G-est      & $1.24$ & $1.07$  & $0.32$ & $0.49$ & $0.39$ & $0.33$ & $0.10$ & $0.14$& $0.03$ & $0.12$ & $0.03$ & $0.04$\\ 
integrate dWOLS & $1.10$ & $0.97$  & $0.31$ & $0.46$ & $0.31$ & $0.19$ & $0.10$ & $0.13$& $0.03$ & $0.13$ & $0.03$ & $0.04$\\ 
integrate G-est & $1.10$ & $0.97$  & $0.31$ & $0.46$ & $0.31$ & $0.19$ & $0.10$ & $0.13$& $0.03$ & $0.13$ & $0.03$ & $0.04$\\ 
CE dWOLS        & $1.18$ & $1.08$  & $0.32$ & $0.48$ & $0.39$ & $0.39$ & $0.10$ & $0.14$& $0.02$ & $0.11$ & $0.03$ & $0.04$\\ 
CE G-est        & $1.18$ & $1.07$  & $0.32$ & $0.48$ & $0.39$ & $0.39$ & $0.10$ & $0.14$& $0.02$ & $0.11$ & $0.03$ & $0.04$\\ 
\hline
\multicolumn{13}{c}{Scenario 2: incorrectly specified treatment and correctly specified treatment-free models} \\
\hline
dWOLS           & $4.58$ & $28.6$ & $0.33$ & $0.53$    & $3.21$ & $26.7$ & $0.10$ & $0.16$   & $2.89$ & $26.8$ & $0.03$ & $0.05$  \\ 
G-est           & $5.57$ & $31.6$ & $0.33$ & $0.53$    & $4.32$ & $29.9$ & $0.10$ & $0.16$   & $4.01$ & $30.1$ & $0.03$ & $0.05$  \\ 
IPTW+dWOLS      & $0.79$ & $7.45$ & $0.33$ & $0.53$    & $$-$0.18$ & $6.06$ & $0.10$ & $0.16$& $$-$0.59$ & $5.96$ & $0.03$ & $0.05$\\ 
IPTW+G-est      & $1.64$ & $11.0$ & $0.33$ & $0.53$    & $0.81$ & $9.64$ & $0.10$ & $0.16$   & $0.41$ & $9.60$ & $0.03$ & $0.05$   \\ 
integrate dWOLS & $1.07$ & $$-$0.01$ & $0.32$ & $0.49$ & $0.31$ & $0.10$ & $0.10$ & $0.15$   & $0.03$ & $0.18$ & $0.03$ & $0.05$   \\ 
integrate G-est & $1.13$ & $0.21$ & $0.32$ & $0.49$    & $0.32$ & $0.13$ & $0.10$ & $0.15$   & $0.04$ & $0.20$ & $0.03$ & $0.05$   \\ 
CE dWOLS        & $1.14$ & $0.07$ & $0.32$ & $0.51$    & $0.39$ & $0.30$ & $0.10$ & $0.15$   & $0.02$ & $0.16$ & $0.03$ & $0.05$\\ 
CE G-est        & $1.20$ & $0.29$ & $0.32$ & $0.52$    & $0.40$ & $0.33$ & $0.10$ & $0.15$   & $0.03$ & $0.18$ & $0.03$ & $0.05$\\ 
\hline
\multicolumn{13}{c}{Scenario 3: correctly specified treatment and incorrectly specified treatment-free models} \\
\hline
dWOLS     & $3.06$ & $10.3$ & $0.33$ & $0.49$  & $2.15$ & $9.28$ & $0.10$ & $0.14$ &  $1.89$ & $9.12$ & $0.03$ & $0.04$ \\ 
G-est     & $3.01$ & $10.2$ & $0.33$ & $0.49$  & $2.13$ & $9.24$ & $0.10$ & $0.14$ &  $1.90$ & $9.13$ & $0.03$ & $0.04$ \\ 
IPTW+dWOLS      & $1.06$ & $0.77$ & $0.32$ & $0.49$  & $0.34$ & $0.36$ & $0.10$ & $0.14$ &  $0.04$ & $0.10$ & $0.03$ & $0.04$ \\ 
IPTW+G-est      & $1.01$ & $0.69$ & $0.32$ & $0.49$  & $0.33$ & $0.31$ & $0.10$ & $0.14$ &  $0.04$ & $0.11$ & $0.03$ & $0.04$ \\ 
integrate dWOLS & $0.84$ & $0.49$ & $0.31$ & $0.46$  & $0.25$ & $0.18$ & $0.10$ & $0.13$ &  $0.05$ & $0.12$ & $0.03$ & $0.04$ \\ 
integrate G-est & $0.86$ & $0.58$ & $0.31$ & $0.46$  & $0.24$ & $0.18$ & $0.10$ & $0.13$ &  $0.05$ & $0.12$ & $0.03$ & $0.04$ \\ 
CE dWOLS        & $0.93$ & $0.62$ & $0.32$ & $0.48$  & $0.33$ & $0.38$ & $0.10$ & $0.14$ &  $0.04$ & $0.10$ & $0.03$ & $0.04$ \\ 
CE G-est        & $0.95$ & $0.70$ & $0.32$ & $0.48$  & $0.32$ & $0.38$ & $0.10$ & $0.14$ &  $0.04$ & $0.10$ & $0.03$ & $0.04$ \\ 
\hline
\end{tabular}} 
\end{table}

\begin{table}[!htbp] \centering 
  \caption{Proportion of the observations for which the optimal partially adaptive treatment strategy is correctly identified across replications} 
  \label{tab4}
		\resizebox{\textwidth}{!}{	
\begin{tabular}{@{\extracolsep{5pt}} lccccccccc} 
\\[-1.8ex]\hline 
\hline \\[-1.8ex] 
& \multicolumn{3}{c}{Scenario 1} & \multicolumn{3}{c}{Scenario 2} & \multicolumn{3}{c}{Scenario 3} \\
n & 100 & 1000 & 10~000 & 100 & 1000 & 10~000 & 100 & 1000 & 10~000 \\ 
\hline \\[-1.8ex] 
dWOLS           & $85.5$ & $98.2$ & $100$ & $75.8$ & $79.5$ & $84.0$ & $85.6$ & $97.9$ & $100$ \\ 
G-est           & $85.5$ & $98.1$ & $100$ & $75.0$ & $76.2$ & $74.8$ & $85.7$ & $98.0$ & $100$ \\ 
IPTW+dWOLS      & $88.2$ & $99.8$ & $100$ & $84.4$ & $97.2$ & $100$ & $88.3$ & $99.8$ & $100$ \\ 
IPTW+G-est      & $88.2$ & $99.8$ & $100$ & $82.8$ & $95.9$ & $100$ & $88.1$ & $99.8$ & $100$ \\ 
integrate dWOLS & $88.8$ & $99.9$ & $100$ & $88.2$ & $99.2$ & $100$ & $89.1$ & $99.8$ & $100$ \\ 
integrate G-est & $88.9$ & $99.9$ & $100$ & $87.9$ & $99.2$ & $100$ & $88.9$ & $99.8$ & $100$ \\ 
CE dWOLS        & $88.4$ & $99.8$ & $100$ & $87.6$ & $99.0$ & $100$ & $88.4$ & $99.7$ & $100$ \\ 
CE G-est        & $88.5$ & $99.8$ & $100$ & $87.6$ & $99.0$ & $100$ & $88.6$ & $99.7$ & $100$ \\ 
\hline \\[-1.8ex] 
\end{tabular}} 
\end{table} 

\clearpage

\section{Application} \label{s:application}

\subsection{Context}
Breast cancer is the most common cancer among women and also the leading cause of cancer deaths among them (Bray et al, 2018). Various treatments can be used to treat patients who receive a breast cancer diagnosis including surgery, chemotherapy, radiotherapy and hormonal therapy. Hormonal therapy seeks stop hormone production or to interfere with the ability of hormones to attach to cancer cells and thus prevent their growth. This treatment is currently recommended for patients whose tumors are hormone receptor positive. Unfortunately, not all cancers that are hormone receptor positive respond adequately to hormonal therapy. Even when hormonal therapy helps treating the cancer itself, it can have multiple undesirable side effects, including hot flashes, sexual problems, weight gain, nausea, fatigue, high cholesterol and osteoporosis (Canadian Cancer Society, 2021). As such, it is important to appropriately tailor hormonal therapy so that it is prescribed only to patients who will benefit from it. 

Obesity is associated with poorer outcomes among patients with breast cancer (Chan et al, 2014). It is also known to affect the circulating levels of estrogen in the body (Calle and Kaaks, 2004). We thus hypothesized that obesity may modify the effect of hormonal therapy on survival in breast cancer patients. More precisely, we believed that hormonal therapy would only be beneficial in hormone receptor positive women with normal weight.  While a breast cancer is generally considered to be hormone receptor positive if it is either estrogen receptor positive (ER+), progesterone receptor positive (PR+) or both, the relevance of testing PR is somewhat controversial. Indeed, it is already recommended to provide hormonal therapy when the cancer is ER+, regardless of PR status, and the profile ER-PR+ is very rare, representing only 3\% of all breast cancers (Dunnwald, Rossing and Li, 2007). In fact, this profile is so rare that some have argued that testing PR has very little therapeutic impact and is unlikely to be cost-effective (Olivotto et al, 2004). Consequently, while effect modification by PR status is expected, tailoring hormonal therapy treatment according to PR status may not be necessary. To illustrate our method, we thus consider the estimation of a single-stage PATS that tailor hormonal therapy decisions according to both ER status and obesity, while neglecting the known interaction with PR status. 

\subsection{Data and analysis}
We analysed data concerning women diagnosed with a non-metastatic breast cancer between 1987 and 2009 that are part of the breast cancer registry maintained by the \textit{Centre des Maladies du Sein}, in Qu\'ebec, Canada. We compared the (log transformed) number of years of survival since breast cancer diagnosis ($Y$) between women who received hormonal therapy ($A = 1$) and those who did not ($A = 0$). Data on age, body mass index (BMI), menopause, cancer's grade and stage, ER status, PR status, type of surgery, first degree familial history of breast cancer, hormone replacement therapy, chemotherapy, radiotherapy, trastuzumab and year of diagnosis were considered as potential confounders ($X$). While ER status, PR status and BMI were all considered as potential effect modifiers ($H$), only ER status and BMI were used as tailoring variables ($H^*$). 

Before proceeding with the analysis, we treated the missing data as follows. When menopause status was missing, it was either imputed to ``no'' when age $<$ 50 and to ``yes'' when age $\geq$ 50. Based on clinical and contextual knowledge, missing data on chemotherapy and radiotherapy were imputed to their most likely value (``no'' for chemotherapy and ``yes'' for radiotherapy). Missing data on progesterone status and cancer's grade were considered as an ``unknown'' category. Observations with missing data on other variables were deleted. A total of 5444 individuals were thus included in the analysis.

To analyze the data, we used both the naive dWOLS estimator and the CE dWOLS estimator we have introduced. For both estimators, the treatment model was fitted using a logistic regression of the treatment according to all potential confounders. The outcome model was a linear regression of the natural logarithm of the survival time according to all potential confounders and including interaction terms between treatment and ER-BMI categories (ER+ and BMI$<$25, ER+ and BMI$\geq$25, ER- and BMI$<$25, ER- and BMI$\geq$25). The results are of this regression model are interpreted as differences in expected log years of survival. Except for years, which was modeled using a restricted cubic splines with five knots, all other covariates were entered in the same categories as presented in Table \ref{TabAppli}. Because the survival time was right censored for 4151 patients, the outcome model was fitted only among observations that were not censored using an inverse probability of censoring weight as proposed by Simoneau et al (2020). Again, the censoring model was a logistic regression including all potential confounders. Censoring weights were truncated at their 99.9$^{th}$ percentile to reduce influence of large weights and to achieve a better representativeness of the weighted uncensored observations to the complete sample. Standard nonparametric bootstrap with 5000 replicates was used to produce confidence intervals.

 \begin{table}[!htbp] \centering 
  \caption{Descriptive statistics of women diagnosed with non-metastatic breast cancer between 1987 and 2010 in the breast cancer registry maintained by the Centre des Maladies du Sein Desch\^enes-Fabia, Qu\'ebec, Canada, according to hormonal therapy use as a treatment for breast cancer} 
  \label{TabAppli} 
\begin{tabular}{@{\extracolsep{5pt}} lrrr} 
\\[-1.8ex]\hline 
\hline \\[-1.8ex] 
                            & No hormonal therapy & Hormonal therapy & SMD \\ 
                            &    1800 (33.0)  &    3644 (67.0)  \\   
\hline \\[-1.8ex] 
Age                                                 &     &     &  0.419 \\ 
\ \ \ $\leq$ 39             &     186 (10.3)  &     119 ( 3.3)  &  \\ 
\ \ \ 40--49                &     496 (27.6)  &     625 (17.2)  &  \\ 
\ \ \ 50--59                &     477 (26.5)  &    1189 (32.6)  &  \\ 
\ \ \ 60--69                &     353 (19.6)  &    1009 (27.7)  &  \\ 
\ \ \ $\geq$ 70             &     288 (16.0)  &     702 (19.3)  &  \\ 
BMI $\geq 25$               &     698 (38.8)  &    1698 (46.6)  &  0.159 \\ 
Menopause                   &     658 (36.5)  &    875  (24.0)  &  0.276 \\ 
Grade                                               &     &     &  0.791 \\ 
\ \ \ 1                     &     225 (12.5)  &    1055 (29.0)  &  \\ 
\ \ \ 2                     &     343 (19.1)  &    1406 (38.6)  &  \\ 
\ \ \ 3                     &     994 (55.2)  &     838 (23.0)  &  \\ 
Unknown                     &     238 (13.2)  &     345 ( 9.5)  &  \\ 
Stage                                               &     &     &  0.077 \\ 
\ \ \ I                     &     809 (44.9)  &    1671 (45.9)  &  \\ 
\ \ \ II                    &     746 (41.4)  &    1569 (43.1)  &  \\ 
\ \ \ III                   &     245 (13.6)  &     404 (11.1)  &  \\ 
Estrogen +                  &     819 (45.5)  &    3517 (96.5)  &  1.360 \\ 
Progesterone                                        &     &     &  0.840 \\ 
\ \ \ +                     &     604 (33.6)  &    2627 (72.1)  &  \\ 
\ \ \ -                     &     999 (55.5)  &     805 (22.1)  &  \\ 
\ \ \ Unknown               &     197 (10.9)  &     212 ( 5.8)  &  \\ 
Surgery                                             &     &     &  0.076 \\ 
\ \ \ Mastectomy            &     482 (26.8)  &     867 (23.8)  &  \\ 
\ \ \ Breast-conserving     &    1289 (71.6)  &    2730 (74.9)  &  \\ 
\ \ \ None                  &      29 ( 1.6)  &      47 ( 1.3)  &  \\ 
1st degree familial history &     399 (22.2)  &     980 (26.9)  &  0.110 \\ 
Hormone replacement therapy &    1170 (65.0)  &    1920 (52.7)  &  0.252 \\ 
Chemotherapy                &     977 (54.3)  &    1527 (41.9)  &  0.250 \\ 
Radiotherapy                &    1307 (72.6)  &    2875 (78.9)  &  0.147 \\ 
Trastuzumab                 &      70 ( 3.9)  &     113 ( 3.1)  &  0.043 \\ 
Year                                                &     &     &  0.729 \\ 
\ \ \ 1985--1989            &     269 (14.9)  &     119 ( 3.3)  &  \\ 
\ \ \ 1990--1994            &     384 (21.3)  &     328 ( 9.0)  &  \\ 
\ \ \ 1995--1999            &     402 (22.3)  &     513 (14.1)  &  \\ 
\ \ \ 2000--2004            &     328 (18.2)  &    1164 (31.9)  &  \\ 
\ \ \ 2005--2009            &     417 (23.2)  &    1520 (41.7)  &  \\ 
\hline \\ [-1.8ex] \\
\end{tabular}
\\All numbers are $n$ (\%). SMD = standardized mean difference.  
\end{table} 

\begin{table}[!htbp] \centering 
\caption{Estimated difference in log survival time in years between women receiving hormonal therapy and those not receiving hormonal therapy according to estrogen receptor (ER) status and body-mass index (BMI)} 
  \label{TabAppli2} 
\resizebox{\textwidth}{!}{
\begin{tabular}{@{\extracolsep{5pt}} lrrrr} 
\\[-1.8ex]\hline 
\hline \\[-1.8ex] 
Estimator & ER+, BMI$<$25 & ER+, BMI$\geq$25 & ER-, BMI$<$25 & ER-, BMI$\geq$25 \\ 
\hline \\[-1.8ex] 
Naive dWOLS & -0.04 (-0.17, 0.11) & 0.12 (-0.12, 0.36) & -0.28 (-0.67, 0.09) & -0.46 (-0.82, -0.09) \\
CE dWOLS   & -0.01 (-0.14, 0.13) & 0.13 (-0.10, 0.34) & -0.28 (-0.68, 0.10) & -0.49 (-0.83, -0.12) \\
\hline \\[-1.8ex] 
\end{tabular}}
\end{table} 

\subsection{Results}

Table \ref{TabAppli} presents the characteristics of the 3644 (67.0\%) patients who received and 1800 (33.0\%) who did not receive hormonal therapy. Among others, women who received hormonal therapy were overall older, had a greater BMI and a higher cancer's grade. Unsurprisingly, there is a large imbalance between treated and untreated patients according to ER and PR statuses, since these markers are used for tailoring treatment in clinical practice. Hormonal therapy has also become more frequent in time. 

Results of the PATS analysis are presented in Table \ref{TabAppli2}. In this specific application, only minor differences were observed between the results of the naive dWOLS and CE dWOLS estimators. In both analyses, hormonal therapy was associated with a lower survival time among ER- patients with a BMI $\geq 25$ (CE dWOLS: -0.49; 95\% CI: -0.83 to -0.12). Among the ER- participants with BMI$<$25, a negative association was also observed, but the results were also compatible with an absence of difference (CE dWOLS: -0.28, 95\%CI: -0.68 to 0.10). Among ER+ participants with BMI $<25$, the observed difference in log survival times was close to zero with 95\% confidence intervals covering equally small beneficial and detrimental associations (CE dWOLS: -0.01, 95\% CI: -0.14 to 0.13). Finally, among ER+ participants with BMI $\geq$ 25, a small positive association was observed, but the confidence intervals indicated the data were compatible with a range of associations going from a slightly negative to moderately positive (CE dWOLS: 0.13, 95\% CI: -0.10 to 0.34).

\section{Discussion} \label{s:discussion}

Better evidence-based tailoring of treatment decisions according to patients' characteristics and to their evolving condition has become a priority. Robust statistical methods for identifying optimal treatment strategies are important tools for reaching this objective. Although it may be theoretically interesting to find the optimal treatment strategy according to multiple, or even all, relevant characteristics, this is often too ambitious or impractical. Indeed, achieving this would require infeasibly large datasets and may render treatment decisions overly complicated. Identifying optimal partially adaptive treatment strategies that aim to tailor treatment decisions according to a few selected characteristics may be a more reasonable goal in practice. 

In this paper, we proposed methods for identifying optimal PATSs that benefit from the double robustness property. Methods for producing inferences were additionally discussed. To facilitate the use of these methods, we supply functions in \texttt{R} as supplementary material. \textcolor{black}{The simulation study we conducted illustrated the double robustness of the estimators we introduced (integrate dWOLS, integrate G-estimation, CE dWOLS and CE G-estimation) and their benefit as compared to alternative estimators of PATS (IPTW+dWOLS or IPTW+G-estimation) or to using ATS estimators when neglecting some effect modifiers.} All our proposed estimators had similar performance in the simulation study in scenarios we investigated. As such, because CE dWOLS is simpler to implement, it may be preferred in practice. The simulation study results also supports the validity of the data-adaptive $m$-out-of-$n$ bootstrap for producing inferences. It is noteworthy that the algorithm for implementing the $m$-out-of-$n$ bootstrap for ATSs (or PATSs) has been developed for the two time-point setting. Extending its application to a multiple time-point setting isn't trivial and should be explored in future studies. Furthermore, while double robustness is an interesting property that offers two chances at correctly specifying models to control confounding bias, it may often be the case in practice that both models are incorrectly specified. There are some situations where double robust methods perform worse than non double robust ones when both models are incorrectly specified, particularly in settings where positivity may be violated (Kang and Schafer, 2007).  

We have also illustrated the usefulness of our method for estimating an optimal PATS concerning the use of hormonal therapy in treating breast cancer patients. In this illustration, we were interested in tailoring treatment only according to ER status and BMI, but not PR status, although effect modification by PR is expected. Neglecting PR status was motivated by the fact that the additional clinical value of PR, above ER alone, is controversial. Although similar results were obtained using our proposed PATS estimator and a naive ATS estimator in this example, this needs not always be the case as illustrated in our simulation study. We believe the lack of difference between ATS and PATS estimators in this specific application might be due to the very low prevalence of the ER-PR+ status. From a substantive point of view, hormonal therapy was associated with poorer survival among ER- women, especially those with a greater BMI. The results were more imprecise regarding ER+ breast cancer patients, but provide some evidence suggesting that ER+ breast cancer patients with a greater BMI may benefit more from hormonal therapy than ER+ patients with a lower BMI. Given the large widths of confidence intervals, more studies are required to further explore this unexpected result. It is possible that residual confounding or selection bias is present although a very rich set of covariates were adjusted for. Moreover, in an analysis where only effect modification according to ER status was considered (not presented), the results were in the expected direction: hormonal therapy was associated with a better survival among ER+ participants, but not among ER- participants. 

In addition to its practical importance, we believe the current work also opens the door to further important methodological developments. First, it seems reasonable to assume that tailoring treatment decisions according to only a few variables may increase the precision of the estimators of the selected interaction terms, since fewer treatment-covariate interaction parameters need to be estimated. It would be worthy to verify this hypothesis and investigate if this reduces the risk of identifying as ``optimal'' treatment strategies that truly have low expected value. PATS estimators could also provide a framework for data-driven selection of the most important tailoring features.

\section*{Acknowledgements}

This work was supported by grants from the Natural Sciences and Engineering Research Council of Canada. DT is also supported by a career award from the Fonds de recherche du Qu\'ebec - Sant\'e. Part of this work was conducted while DT was a visiting professor at McGill University. DT would like thank Université Laval for funding his visit at McGill University, and McGill University for welcoming him.  \vspace*{-8pt}

\bibliographystyle{biorefs}

\clearpage
\section*{}

{\center \huge  Supplementary material for ``Double robust estimation of partially adaptive treatment strategies'' by Denis Talbot, Erica EM Moodie and Caroline Diorio. }

\section*{Appendix A - G-estimation of adaptive treatment strategies}

The G-estimation of the parameters $\psi_j$ is performed recursively, for $j = K, ..., 1$. First, a vector-valued function $S(A_j)$ of the same length as $\psi_j$ containing effect modifiers of the treatment effect must be specified. The optimal form for $S(A_j)$ is known (Robins, 2004), but is generally complex and $S(a_j) = \partial \gamma_j(a_j, h_j; \psi_j)/\partial \psi_j$ is thus commonly used. For example, if $\gamma_j(a_j, h_j; \psi_j) = a_j(\psi_{0j} + \psi_{1j}h_j)$ then $S(a_j)$ could be $(a_j, a_j h_j)$. Next, a model for $\mathbb{E}[S(A_j)|H_j]$ is specified. Define $G_{iK} = Y_i - \gamma_K(a_{iK}, h_{iK}; \psi_K)$ and $G_{ij} = Y_i - \gamma_j(a_{ij}, h_{ij}; \psi_j) + \sum_{k = j+1}^K [\gamma_k(d_k^{opt}, h_{ik}; \hat{\psi}_k) - \gamma_k(a_{ik}, h_{ik}; \hat{\psi}_k)]$ if $j \neq K$, where $\hat{\psi}_k$ is the estimated value of $\psi_k$. A model for $\mathbb{E}[G_j|H_j]$ is then specified. The G-estimate of $\psi_j$ is the solution $\hat{\psi}_j$ to the estimating equations $0 = \sum_{i=1}^n \{S(A_{ij}) - \mathbb{E}[S(A_{ij})|H_{ij}]\}\{G_{ij} - \mathbb{E}[G_{ij}|H_{ij}]\}$. This equation has a closed-form solution for continuous $Y$ when linear models are used for the treatment-free $(\mathbb{E}[G_{ij}|H_{ij}])$ and blip components.

\section*{Appendix B - Double robustness of weighted ordinary least squares}

\textcolor{black}{We provide a proof of the double robustness of the dWOLS estimator of the parameters of the blip $\psi$ in a single time-point setting. As such, we will use a simplified notation, dropping the subscript $j$ and noting that $H_1 = X_1$. A  sketch proof for the multiple time-point setting is then provided.}

Consider the case where the outcome model may be misspecified, but the treatment model is correctly specified. We consider the weights $w = |A - \mathbb{E}[A|X]|$, which benefit from the balancing property $A\coprod X$ in the weighted data. \textcolor{black}{The proof would proceed similarly for other types of balancing weights.} We show that $\mathbb{E}[\hat{\psi}]$ converges in probability to $\psi$. We first consider the following outcome model, for which the treatment-free component ($X_i \beta$) may be misspecified: $Y_i = X_i \beta + A_i X_i \psi + \epsilon_i$. \textcolor{black}{We note that instead of representing the variables themselves, $X$ can be taken as a design matrix that features transformation of the variables, such as quadratic terms or covariate-covariate interactions. As such, this outcome model specification is fairly general.} The weighted least squares estimating equations are\\
\begin{align*}
0 &= \sum w_i X^\top_i (Y_i - X_i \hat{\beta} - A_i X_i \hat{\psi}) \\
0 &= \sum w_i A_i X^\top_i (Y_i - X_i \hat{\beta} - A_i X_i \hat{\psi}).
\end{align*}
Isolating $\hat{\beta}$ in both equations
\begin{align*}
\hat{\beta} &= (\sum w_i X_i^\top X_i)^{-1}[(\sum w_i X_i^\top Y_i) - (\sum w_i A_i X_i^\top X_i)\psi] \\
\hat{\beta} &= (\sum w_i A_i X_i^\top X_i)^{-1}[(\sum w_i A_i X_i^\top Y_i) - (\sum w_i A_i X_i^\top X_i)\psi], 
\end{align*}
then equating both equations, the estimator for $\psi$ is obtained
\begin{align*}
\hat{\psi} &= [(\sum w_i X_i^\top X_i)^{-1} (\sum w_i A_i X_i^\top X_i) - I_p]^{-1} \times \\
& [(\sum w_i X_i^\top X_i)^{-1}(\sum w_i X_i^\top Y_i) - (\sum w_i A_i X_i^\top X_i)^{-1}(\sum w_i A_i X_i^\top Y_i)] \\
& = [(\sum w_i (A_i + (1 - A_i)) X_i^\top X_i)^{-1} (\sum w_i A_i X_i^\top X_i) - I_p]^{-1} \times \\
& [(\sum (A_i + (1 - A_i)) w_i X_i^\top X_i)^{-1}(\sum w_i X_i^\top Y_i) - (\sum w_i A_i X_i^\top X_i)^{-1}(\sum w_i A_i X_i^\top Y_i)]. \\
\end{align*}
Since $\mathbb{E}[wA\eta(X)] = \mathbb{E}[w(1 - A)\eta(X)] = \mathbb{E}[\pi(x)\eta(X)]$, where $\eta(X)$ is any function of $X$ and $\pi(X) = P(A=1|X)P(A=0|X)$ (Li and Li 2019), we get
\begin{align*}
\hat{\psi} &\stackrel{p}{\rightarrow} [(2\sum \pi(X_i) X_i^\top X_i)^{-1} (\sum \pi(X_i) X_i^\top X_i) - I_p]^{-1} \times \\
& [(2\sum \pi(X_i) X_i^\top X_i)^{-1}(\sum w_i X_i^\top Y_i) - (\sum \pi(X_i) X_i^\top X_i)^{-1}(\sum w_i A_i X_i^\top Y_i)] \\
&= -[(\sum \pi(X_i) X_i^\top X_i)^{-1}(\sum w_i X_i^\top Y_i) - 2(\sum \pi(X_i) X_i^\top X_i)^{-1}(\sum w_i A_i X_i^\top Y_i)] \\
&= -(\sum \pi(X_i) X_i^\top X_i)^{-1}[\sum (w_i X_i^\top Y_i - 2 w_i A_i X_i^\top Y_i)] \\
&= -(\sum \pi(X_i) X_i^\top X_i)^{-1}[\sum (w_i (A_i + (1 - A_i)) X_i^\top Y_i - 2 w_i A_i X_i^\top Y_i)] \\
&= (\sum \pi(X_i) X_i^\top X_i)^{-1}[\sum w_i A_i X_i^\top Y_i - w_i (1 - A_i) X_i^\top Y_i)] \\
\end{align*}
Taking the expectation on both sides
\begin{align}
\mathbb{E}\left[\hat{\psi}\right] &\stackrel{p}{\rightarrow} 
\mathbb{E}\left\{(\sum \pi(X_i) X_i^\top X_i)^{-1}[\sum w_i A_i X_i^\top Y_i - w_i (1 - A_i) X_i^\top Y_i]\right\} \nonumber \\
&= \mathbb{E}\left[\mathbb{E}\left\{(\sum \pi(X_i) X_i^\top X_i)^{-1}[\sum w_i A_i X_i^\top Y_i - w_i (1 - A_i) X_i^\top Y_i]|X_i \right\} \right] \nonumber  \\
&= \mathbb{E}\left[(\sum \pi(X_i) X_i^\top X_i)^{-1} \sum X_i^\top \left\{\mathbb{E}\left[w_i A_i Y_i|X_i\right] - \mathbb{E}\left[w_i(1 - A_i) Y_i|X_i\right]\right\}\right]. \label{eq1}
\end{align}
Consider $\mathbb{E}\left[w_i A_i Y_i|X_i\right]$:
\begin{align*}
\mathbb{E}\left[w_i A_i Y_i|X_i\right] &= \mathbb{E}\left[w_i Y_i^1|A_i = 1, X_i\right]P(A_i = 1|X_i) \\
&= \mathbb{E}[w_i|A_i = 1, X_i] \mathbb{E}\left[Y_i^1|X_i\right]P(A_i = 1|X_i) \\
&= P(A_i = 0|X_i) \mathbb{E}\left[Y_i^1|X_i\right]P(A_i = 1|X_i) \\
& = \pi(X_i) \mathbb{E}\left[Y_i^1|X_i\right],
\end{align*}
where the first equality is obtained using the consistency assumption and the second using the exchangeability assumption. Similarly, we can show that $\mathbb{E}\left[w_i A_i Y_i|X_i\right] = \pi(X_i)\mathbb{E}\left[Y_i^0|X_i\right]$. Inserting these results in (\ref{eq1})
\begin{align*}
\mathbb{E}\left[\hat{\psi}\right] &\stackrel{p}{\rightarrow} 
 \mathbb{E}\left\{(\sum \pi(X_i) X_i^\top X_i^\top)^{-1} \sum \pi(X_i) X_i^\top \mathbb{E}\left[Y_i^1 - Y_i^0|X_i\right]\right\} \\
&= \mathbb{E}\left\{(\sum \pi(X_i) X_i^\top X_i^\top)^{-1} \sum \pi(X_i) X_i^\top X_i \psi \right\} = \psi. 
\end{align*}

Now consider the case where the outcome model is correctly specified, but the treatment model may be misspecified. Denote $Z = (X, AX)$ and $\theta = (\beta, \psi)^\top$, we now show that $\mathbb{E}[\hat{\theta}] = \theta$. 
\begin{align*}
\mathbb{E}[\hat{\theta}] &= \mathbb{E}\left[(Z^\top W Z)^{-1} Z^\top W Y \right] \\
&= \mathbb{E}\left[(Z^\top W Z)^{-1} Z^\top W (Z\theta + \varepsilon) \right] \\
&= \mathbb{E}\left[(Z^\top W Z)^{-1} X^\top W Z\theta + (Z^\top W Z)^{-1} Z^\top W \varepsilon \right] \\
&= \theta + \mathbb{E}\left[(Z^\top W Z)^{-1} Z^\top W \varepsilon \right] \\
&= \theta
\end{align*}
\noindent because $Z^\top W \varepsilon = 0$ by construction. 

In the multiple time-point setting, we can first note that the consistency of the dWOLS estimator for the parameters of the blip at the last time-point ($j = K$) follows directly from the previous results. For $j < K$, we assume that the later blips were estimated consistently. As a consequence, ${\mathbb{E}[\tilde{Y}_j] = \mathbb{E}\{Y + \sum_{k=j+1}^K[\gamma_k(d_k^{opt}, h_k; \hat{\psi}_k) - \gamma_k(a_{ik}, h_k; \hat{\psi}_k)]\}}$ is consistent for ${\mathbb{E}\{Y + \sum_{k=j+1}^K[\gamma_k(d_k^{opt}, h_k; \psi_k) - \gamma_k(a_{ik}, h_k; \psi_k)]\}}$. Under our causal assumptions, it can be verified that $\mathbb{E}\{Y + \sum_{k=j+1}^K[\gamma_k(d_k^{opt}, h_{k}; \psi_k) - \gamma_k(a_{k}, h_k; \psi_k)]|A_j, H_j\} = \mathbb{E}[Y^{\bar{a}_{j},\underline{d}_{j+1}^{opt}}|H_j]$. Together, these results imply that $\mathbb{E}[\tilde{Y}_j|A_j, H_j]$ is a consistent estimator for $\mathbb{E}[Y^{\bar{a}_{j},\underline{d}_{j+1}^{opt}}|H_j]$. The rest of the proof follows the same steps as above, replacing $Y$ by $\tilde{Y}$, $X$ by $H$ and adding back the subscript $j$. 
\clearpage

\section*{Appendix C - G-estimation estimators of partially adaptive treatment strategies}

This Appendix details the steps of the G-estimation analogues to the dWOLS estimators of PATS presented in the main manuscript.

\subsection*{IPTW+G-estimation}

\begin{flushleft}
\begin{enumerate}
	\item Estimate $\psi_j$ with the G-estimation algorithm.
	\item Specify a vector valued function $S^*(A_j)$ of the same length as $\psi_j^*$, containing the effect modifiers $H_j^*$.
	\item Define $G_{iK}^* = Y_i - \gamma_j^*(a_{iK}, h_{iK}^*; \psi_K^*)$ and ${G_{ij}^* = Y_i - \gamma_j^*(a_{ij}, h_{ij}^*; \psi_j^*) + \sum_{k = j+1}^K [\gamma_k^*(d_k^{opt*}, h_{ik}; \hat{\psi}_k^*) - \gamma_k(a_{ik}, h_{ik}; \hat{\psi}_k)]}$ if $j \neq K$.
	\item Specify a model for $\mathbb{E}[G_j^*|H_j]$.
	\item Specify a model for $\mathbb{E}[S^*(A_j)|H_j^*]$.
	\item Compute weights $\varpi(A_j, H_j, H_j^*) = \frac{P(A_j|H_j^*)}{P(A_j|H_j)}$.
	\item $\hat{\psi}_j^*$ is the solution $\psi_j^*$ to the estimating equations ${0 = \sum_{i=1}^n \varpi(A_{ij}, H_{ij}, H_{ij}^*) \{S^*(A_{ij}) - \mathbb{E}[S^*(A_{ij})|H_{ij}^*]\}\{G_{ij}^* - \mathbb{E}[G_{ij}^*|H_{ij}]\}}$.
\end{enumerate}
\end{flushleft}

\subsection*{Integrate G-estimation}
\begin{flushleft}
\begin{enumerate}
	\item Estimate $\psi_j$ with the G-estimation algorithm.
	\item Specify a vector valued function $S(A_j)$ of the same length as $\psi_j$, containing the effect modifiers among $H_j$.
	\item Define $G_{iK}^* = Y_i - \gamma_j^{\dagger}(a_{iK}, h_{iK})$ and ${G_{ij}^* = Y_i - \gamma_j^{\dagger}(a_{ij}, h_{ij}) + \sum_{k = j+1}^K [\gamma_k^*(d_k^{opt*}, h_{ik}; \hat{\psi}_k^*) - \gamma_k(a_{ik}, h_{ik}; \hat{\psi}_k)]}$ if $j \neq K$.
	\item Specify a model for $\mathbb{E}[G_j^*|H_j]$.
	\item Specify a model for $\mathbb{E}[S(A_j)|H_j]$.
	\item Estimate $\gamma_j^{\dagger}(a_j, h_j)$ as the solution to ${0 = \sum_{i=1}^n \{S(A_{ij}) - \mathbb{E}[S(A_{ij})|H_{ij}]\}\{G_{ij}^* - \mathbb{E}[G_{ij}^*|H_{ij}]\}}$.
	\item Estimate $f_{H_j^C|H_j^*}$ and compute $\hat{\gamma}_j^*(a_j, h_j^*) = \int_{H_j^C} \hat{\gamma_j}^\dagger(a_j, h_j) \hat{f}_{H_j^C|H_j^*} \mu(H_j^C)$.
\end{enumerate}
\end{flushleft}

\subsection*{CE G-estimation}

All steps except the last one are the same as in the algorithm of ``integrate G-estimation.'' This step is replaced by the two following steps

\begin{flushleft}
\begin{enumerate}
  \setcounter{enumi}{6}
	\item Compute $Q(a, h_{ij}) = \hat{\gamma}^{\dagger}(a, h_{ij})$ for all observations.
	\item Run a regression of $\mathbb{E}[Q(a, H_j)|H_j^*] = \gamma^*(a, h_j^*)$.
\end{enumerate}
\end{flushleft}

\clearpage

\section*{Appendix D - Connection between the IPTW+dWOLS and IPTW+G-estimation estimators}

We show that the estimate of the IPTW+dWOLS is also a solution to the IPTW+G-estimation estimating equations in certain circumstances to illustrate the close connection between IPTW+dWOLS and IPTW+G-estimation. Our presentation closely follows the connection between G-estimation and dWOLS for estimating ATSs made by Wallace and Moodie (2015). We consider the single time point setting where $\gamma^*(a, h^*) = A H^*\psi^*$, $\mathbb{E}[G|H^*] = f(h^*; \bm{\beta}) = \bm{\beta} \bm{H}^*$ and $S^*(A) = \partial \gamma_j^*(a_j, h_j^*; \psi_j^*)/\partial \psi_j^* = AH^*$, where $H^*$ is assumed to include a column of 1 for the intercept. The estimating equations of IPTW+G-estimation can be written as:
\begin{equation}
\begin{aligned}
0 = \sum_{a_i = 0}& \varpi(a_i, h_i, h_i^*) h^*_i \left(-\mathbb{E}[A|H^*_{i}]   \right)(y_i - \bm{\beta} h_i) \\
+& \sum_{a_i = 1} \varpi(a_i, h_i, h_i^*) h^*_i \left(1-\mathbb{E}[A|H^*_i]   \right)(y_i - \psi^*h^*_i - \bm{\beta}h_i).
\end{aligned}
\label{eq0}
\end{equation}
The dWOLS estimating equations can be written in a similar form: 
\begin{align*}
0 = \sum_{a_i = 0}& \varpi(a_i, h_i, h_i^*) \times w^*(a_i, h_i^*) h_i (y_i - \bm{\beta} h_i) \\
+& \sum_{a_i = 1} \varpi(a_i, h_i, h_i^*) \times w^*(a_i, h_i^*) h_i (y_i - \psi^*h^*_i - \bm{\beta} h_i)\\
0 = \sum_{a_i =1} & \varpi(a_i, h_i, h_i^*) \times w^*(a_i, h_i^*) h^*_i (y_i - \bm{\psi}^*h^*_i - \bm{\beta} h_i).
\end{align*}
\noindent Let $w(a_i, h_i^*) = |A_i - \mathbb{E}[A|h_i^*]|$. The dWOLS estimating equations are: 
\begin{align}
0 = \sum_{a_i = 0}& \varpi(a_i, h_i, h_i^*) h_i \left(\mathbb{E}[A|h_i^*]   \right) (y_i - \bm{\beta} h_i) \nonumber \\
+& \sum_{a_i = 1}  \varpi(a_i, h_i, h_i^*) h_i \left(1-\mathbb{E}[A|h_i^*]   \right) (y_i - \psi^*h^*_i - \bm{\beta} h_i) \label{eq1.1} \\
0 = \sum_{a_i = 1}& \varpi(a_i, h_i, h_i^*) h^*_i \left(1-\mathbb{E}[A|h_i^*]   \right) (y_i - \psi^*h^*_i - \bm{\beta} h_i). \label{eq2}
\end{align}
Remarking that $h_i = (h_i^*, h_{i}^C)$, then writing the system of equations \{(\ref{eq2}) - (\ref{eq1.1}), (\ref{eq2})\}, we get
\begin{align}
0 & = \sum_{a_i = 0} \varpi(a_i, h_i, h_i^*) h^*_i \left(-\mathbb{E}[A|h_i^*]   \right) (y_i - \bm{\beta} h_i) \label{eq4}\\
0 & = \sum  \varpi(a_i, h_i, h_i^*) h_i^C \left(a_i -\mathbb{E}[A|h_i^*]   \right) (y_i - \bm{\beta} h_i - a_i \bm{\psi}^*h^*_i) \label{eq5}  \\
0 & = \sum_{a_i = 1} \varpi(a_i, h_i, h_i^*) h^*_i \left(1-\mathbb{E}[A|h_i^*]   \right) (y_i - \bm{\psi}^*h^*_i - \bm{\beta} h_i). \label{eq6} 
\end{align}
Since the sum of the first and last equations from this IPTW+dWOLS is equal to the estimating equations from IPTW+G-estimation [(\ref{eq4}) + (\ref{eq6}) = (\ref{eq0})], the solution of IPTW+dWOLS is also a solution to the IPTW+G-estimation estimating equations. 

\clearpage

\section*{Appendix E - Additional simulation results with two time-points}

We consider three additional scenarios with two time-points. At each time-point, two pre-treatment covariates are measured, $X_1$ and $X_2$. Both are confounders and effect modifiers for the treatment at their respective time-point. However, only $X_1$ is intended to be used for tailoring treatment. Using the notation introduced in the paper, we have $H_1 = (X_{11}, X_{12})$, $H_2 = (X_{11}, X_{12}, A_1, X_{21}, X_{22})$, $H_1^* = X_{11}$, $H_2^* = X_{21}$. 

In all scenarios
\begin{align*}
X_{11} \sim Bernoulli(p = 0.5), \\
X_{12} \sim Bernoulli(p = 0.5). \\
\end{align*}
\noindent In Scenarios 1 and 3, $A_1 \sim Bernoulli[p = expit(-1 + X_{11} + X_{12})]$, whereas in Scenario 2, ${A_1 \sim Bernoulli[p = expit(-1 + X_{11} + X_{12} + X_{11} X_{12})]}$. In all scenarios
\begin{align*}
X_{21} \sim Bernoulli[p = expit(-1 + X_{11} + A_1)], \\
X_{22} \sim Bernoulli[p = expit(-1 + X_{12} + A_1)]. \\
\end{align*}
\noindent In Scenarios 1 and 3, $A_2 \sim Bernoulli[p = expit(-1 + X_{21} + X_{22})]$, and in Scenario 2 ${A_2 \sim Bernoulli[p = expit(-1 + X_{21} + X_{22} + X_{21}X_{22})]}$.  In all scenarios
\begin{align*}
A_1^{opt} = I(\bm{X}_1\psi_1 > 0),\\
\mu_1 = (A_1^{opt} - A_1)\bm{X}_1 \psi_1,\\
A_2^{opt} = I(\bm{X}_2\psi_2 > 0),\\
\mu_2 = (A_2^{opt} - A_2)\bm{X}_2 \psi_2,  
\end{align*}
\noindent where $I(\cdot)$ is the usual indicator function, $\bm{X}_j = (1, X_{j1}, X_{j2}), j = 1,2$, and $\psi_1 = \psi_2 = (-0.5, 1, -1)$. Finally, in Scenario 1 and 2, $Y \sim N(X_{11} + X_{12} - \mu_1 - \mu_2, 1)$, and in Scenario 3, $Y \sim N(X_{11} + X_{12} + X_{11}X_{12} - \mu_1 - \mu_2, 1)$. 

The true value of the causal effect at the second time-point is $\gamma_2^*(A_2, X_{12}) = A_2(\psi_{20}^* + \psi_{21}^* X_{21}) = A_2\{(\psi_{20} + \mathbb{E}[X_{22}|X_{21} = 0]\psi_{22}) + [\psi_{21} + (\mathbb{E}[X_{22}|X_{21} = 1] - \mathbb{E}[X_{22}|X_{21} = 0])\psi_{22}]X_{21}\} \approx A_2(-1.01 + 1.03X_{21})$. The true value of the causal effects at the first time-point were estimated through Monte Carlo simulations of the counterfactuals with a sample size of 1~000~000, $\gamma_1^*(A_1, X_{11}) \approx A_1 (-0.96 + 0.91 X_{11})$.

As in the main simulation, the treatment and the treatment-free models only include main terms. Hence, in Scenario~1, both the treatment score and the treatment-free models are correctly specified. In Scenario~2, only the treatment-free model is correctly specified. In Scenario~3, only the treatment model is correctly specified. 

A total of 1000 replications of each scenario were generated for $n = 100, 1000$ and $10~000$. The results are summarized in Tables~1-4. As in the main simulations, standard G-estimation and dWOLS were biased in all three scenarios. In Scenarios~1 and 3, where the treatment model was correctly specified, the PATS estimators we proposed produced estimates with very low bias, especially for larger sample sizes. When the treatment model was misspecified, some bias remained for IPTW+G-estimation and IPTW+dWOLS. The bias for the other methods was somewhat larger when the treatment model was misspecified then when it was correctly specified, but could be considered as negligible (<5\% for n = 100, and <3\% for n = 1000 and n~=~10~000). Standard ATS estimators identified the optimal PATS less often than the PATS estimators under all Scenarios, especially at larger sample sizes. The expected loss when the optimal PATS was incorrectly estimated was similar across estimators and varied from 0.076 to 1.102 with a mean of 0.128. The expected loss was slightly larger in Scenario~2 than in Scenarios~1 and 3 with means varying between $\approx$ 0.13 and 0.17 depending on the estimator and sample size.

\begin{table}[h!] \centering 
  \caption{Results of Scenario 1 (correctly specified treatment and treatment-free models), $n$ = 100 (top), 1000 (middle) and 10~000 (bottom)}
		\resizebox{\textwidth}{!}{	
\begin{tabular}{@{\extracolsep{5pt}} lrrrrrrrrr} 
\\[-1.8ex]\hline 
\hline \\[-1.8ex] 
& \multicolumn{4}{c}{Rel. bias} & \multicolumn{4}{c}{SD} \\
methods & $\psi_{01}$ & $\psi_{11}$ & $\psi_{02}$ & $\psi_{12}$ & $\psi_{01}$ & $\psi_{11}$ & $\psi_{02}$ & $\psi_{12}$  \\ 
\hline \\[-1.8ex] 
dWOLS & $13.697$ & $11.299$ & $4.975$ & $12.779$ & $0.341$ & $0.475$ & $0.340$ & $0.486$ \\ 
G-estimation & $13.511$ & $11.150$ & $4.037$ & $11.546$ & $0.342$ & $0.475$ & $0.343$ & $0.494$ \\ 
IPTW+dWOLS & $$-$1.909$ & $2.862$ & $$-$0.380$ & $1.758$ & $0.340$ & $0.473$ & $0.344$ & $0.485$ \\ 
IPTW+G-estimation  & $$-$2.220$ & $2.792$ & $$-$2.065$ & $$-$0.994$ & $0.340$ & $0.474$ & $0.352$ & $0.501$ \\ 
integrate dWOLS & $$-$1.979$ & $2.361$ & $$-$0.979$ & $0.079$ & $0.335$ & $0.470$ & $0.338$ & $0.484$ \\ 
integrate G-estimation & $$-$2.296$ & $2.324$ & $$-$2.048$ & $$-$1.537$ & $0.336$ & $0.471$ & $0.343$ & $0.496$ \\ 
CE dWOLS & $$-$1.979$ & $2.361$ & $$-$0.979$ & $0.079$ & $0.335$ & $0.470$ & $0.338$ & $0.484$ \\ 
CE G-estimation & $$-$2.296$ & $2.324$ & $$-$2.048$ & $$-$1.537$ & $0.336$ & $0.471$ & $0.343$ & $0.496$ \\ 
\hline \\[-1.8ex] 
dWOLS & $15.808$ & $10.739$ & $6.183$ & $12.698$ & $0.100$ & $0.140$ & $0.102$ & $0.142$ \\ 
G-estimation & $15.799$ & $10.726$ & $6.169$ & $12.652$ & $0.100$ & $0.141$ & $0.101$ & $0.142$ \\ 
IPTW+dWOLS & $$-$0.609$ & $0.426$ & $0.086$ & $$-$0.102$ & $0.102$ & $0.139$ & $0.102$ & $0.142$ \\ 
IPTW+G-estimation & $$-$0.625$ & $0.420$ & $0.021$ & $$-$0.241$ & $0.102$ & $0.139$ & $0.102$ & $0.142$ \\ 
integrate dWOLS & $$-$0.575$ & $0.415$ & $0.039$ & $$-$0.223$ & $0.101$ & $0.139$ & $0.101$ & $0.142$ \\ 
integrate G-estimation & $$-$0.591$ & $0.415$ & $0.018$ & $$-$0.278$ & $0.101$ & $0.139$ & $0.101$ & $0.142$ \\ 
CE dWOLS & $$-$0.575$ & $0.415$ & $0.039$ & $$-$0.223$ & $0.101$ & $0.139$ & $0.101$ & $0.142$ \\ 
CE G-estimation & $$-$0.591$ & $0.415$ & $0.018$ & $$-$0.278$ & $0.101$ & $0.139$ & $0.101$ & $0.142$ \\ 
\hline \\[-1.8ex] 
dWOLS & $16.419$ & $10.305$ & $6.286$ & $12.998$ & $0.032$ & $0.046$ & $0.032$ & $0.044$ \\ 
G-estimation & $16.419$ & $10.304$ & $6.281$ & $12.991$ & $0.032$ & $0.046$ & $0.032$ & $0.044$ \\ 
IPTW+dWOLS & $$-$0.138$ & $0.022$ & $0.064$ & $0.017$ & $0.032$ & $0.045$ & $0.031$ & $0.043$ \\ 
IPTW+G-estimation & $$-$0.140$ & $0.021$ & $0.056$ & $0.001$ & $0.032$ & $0.045$ & $0.031$ & $0.043$ \\ 
integrate dWOLS & $$-$0.135$ & $0.018$ & $0.056$ & $0.009$ & $0.032$ & $0.045$ & $0.031$ & $0.043$ \\ 
integrate G-estimation & $$-$0.136$ & $0.018$ & $0.050$ & $0.001$ & $0.032$ & $0.045$ & $0.031$ & $0.043$ \\ 
CE dWOLS & $$-$0.135$ & $0.018$ & $0.056$ & $0.009$ & $0.032$ & $0.045$ & $0.031$ & $0.043$ \\ 
CE G-estimation & $$-$0.136$ & $0.018$ & $0.050$ & $0.001$ & $0.032$ & $0.045$ & $0.031$ & $0.043$ \\ 
\hline \\[-1.8ex] 
\end{tabular}} 
\end{table} 

\begin{table}[!htbp] \centering 
  \caption{Results of Scenario 2 (incorrectly specified treatment models and correctly specified treatment-free models), $n$ = 100 (top), 1000 (middle) and 10~000 (bottom)} 
		\resizebox{\textwidth}{!}{	
\begin{tabular}{@{\extracolsep{5pt}} lrrrrrrrrr} 
\\[-1.8ex]\hline 
\hline \\[-1.8ex] 
& \multicolumn{4}{c}{Rel. bias} & \multicolumn{4}{c}{SD} \\
methods & $\psi_{01}$ & $\psi_{11}$ & $\psi_{02}$ & $\psi_{12}$ & $\psi_{01}$ & $\psi_{11}$ & $\psi_{02}$ & $\psi_{12}$  \\ 
\hline \\[-1.8ex] 
dWOLS & $15.459$ & $25.542$ & $4.349$ & $30.844$ & $0.344$ & $0.508$ & $0.355$ & $0.549$ \\ 
G-estimation & $15.147$ & $25.029$ & $3.210$ & $29.522$ & $0.344$ & $0.507$ & $0.356$ & $0.554$ \\ 
IPTW+dWOLS & $$-$6.027$ & $8.677$ & $$-$3.522$ & $10.824$ & $0.349$ & $0.526$ & $0.358$ & $0.557$ \\ 
IPTW+G-estimation & $$-$7.315$ & $8.553$ & $$-$6.127$ & $6.771$ & $0.354$ & $0.526$ & $0.369$ & $0.579$ \\ 
integrate dWOLS & $$-$3.752$ & $4.788$ & $$-$3.129$ & $3.718$ & $0.341$ & $0.510$ & $0.353$ & $0.553$ \\ 
integrate G-estimation & $$-$5.007$ & $3.558$ & $$-$4.132$ & $$-$0.778$ & $0.345$ & $0.513$ & $0.359$ & $0.572$ \\ 
CE dWOLS & $$-$3.752$ & $4.788$ & $$-$3.129$ & $3.718$ & $0.341$ & $0.510$ & $0.353$ & $0.553$ \\ 
CE G-estimation & $$-$5.007$ & $3.457$ & $$-$4.533$ & $0.053$ & $0.345$ & $0.513$ & $0.359$ & $0.572$ \\ 
\hline \\[-1.8ex] 
dWOLS & $16.909$ & $24.029$ & $5.388$ & $28.569$ & $0.102$ & $0.154$ & $0.104$ & $0.156$ \\ 
G-estimation & $16.950$ & $24.108$ & $5.474$ & $29.050$ & $0.102$ & $0.154$ & $0.104$ & $0.156$ \\ 
IPTW+dWOLS & $$-$4.500$ & $4.120$ & $$-$3.546$ & $3.408$ & $0.103$ & $0.152$ & $0.105$ & $0.160$ \\ 
IPTW+G-estimation & $$-$5.017$ & $4.589$ & $$-$3.802$ & $3.956$ & $0.104$ & $0.152$ & $0.105$ & $0.160$ \\ 
integrate dWOLS & $$-$2.520$ & $1.870$ & $$-$2.279$ & $0.970$ & $0.103$ & $0.151$ & $0.104$ & $0.158$ \\ 
integrate G-estimation & $$-$3.025$ & $1.088$ & $$-$2.368$ & $0.110$ & $0.103$ & $0.151$ & $0.104$ & $0.159$ \\ 
CE dWOLS & $$-$2.520$ & $1.870$ & $$-$2.279$ & $0.970$ & $0.103$ & $0.151$ & $0.104$ & $0.158$ \\ 
CE G-estimation & $$-$3.025$ & $1.088$ & $$-$2.368$ & $0.110$ & $0.103$ & $0.151$ & $0.104$ & $0.159$ \\ 
\hline \\[-1.8ex] 
dWOLS & $16.941$ & $23.487$ & $5.425$ & $29.256$ & $0.033$ & $0.050$ & $0.032$ & $0.047$ \\ 
G-estimation & $17.010$ & $23.641$ & $5.541$ & $29.779$ & $0.033$ & $0.050$ & $0.032$ & $0.047$ \\ 
IPTW+dWOLS & $$-$3.724$ & $3.541$ & $$-$3.706$ & $3.686$ & $0.032$ & $0.049$ & $0.032$ & $0.047$ \\ 
IPTW+G-estimation & $$-$4.244$ & $4.067$ & $$-$3.861$ & $4.480$ & $0.033$ & $0.049$ & $0.032$ & $0.047$ \\ 
integrate dWOLS & $$-$1.823$ & $1.438$ & $$-$2.289$ & $1.511$ & $0.033$ & $0.049$ & $0.032$ & $0.046$ \\ 
integrate G-estimation & $$-$2.304$ & $0.716$ & $$-$2.351$ & $0.713$ & $0.033$ & $0.049$ & $0.032$ & $0.046$ \\ 
CE dWOLS & $$-$1.823$ & $1.438$ & $$-$2.289$ & $1.511$ & $0.033$ & $0.049$ & $0.032$ & $0.046$ \\ 
CE G-estimation  & $$-$2.304$ & $0.716$ & $$-$2.351$ & $0.713$ & $0.033$ & $0.049$ & $0.032$ & $0.046$ \\ 
\hline \\[-1.8ex] 
\end{tabular} }
\end{table} 

\begin{table}[!htbp] \centering 
  \caption{Results of Scenario 3 (correctly specified treatment models and incorrectly specified treatment-free models), $n$ = 100 (top), 1000 (middle) and 10~000 (bottom)} 
		\resizebox{\textwidth}{!}{	
\begin{tabular}{@{\extracolsep{5pt}} lrrrrrrrrr} 
\\[-1.8ex]\hline 
\hline \\[-1.8ex] 
& \multicolumn{4}{c}{Rel. bias} & \multicolumn{4}{c}{SD} \\
methods & $\psi_{01}$ & $\psi_{11}$ & $\psi_{02}$ & $\psi_{12}$ & $\psi_{01}$ & $\psi_{11}$ & $\psi_{02}$ & $\psi_{12}$  \\ 
\hline \\[-1.8ex] 
dWOLS & $14.480$ & $12.464$ & $4.935$ & $12.580$ & $0.353$ & $0.478$ & $0.356$ & $0.507$ \\ 
G-estimation & $14.190$ & $12.173$ & $3.927$ & $11.132$ & $0.353$ & $0.477$ & $0.360$ & $0.516$ \\ 
IPTW+dWOLS & $$-$0.830$ & $4.613$ & $$-$0.399$ & $1.504$ & $0.356$ & $0.478$ & $0.360$ & $0.506$ \\ 
IPTW+G-estimation & $$-$1.223$ & $4.447$ & $$-$2.102$ & $$-$1.400$ & $0.356$ & $0.478$ & $0.369$ & $0.524$ \\ 
integrate dWOLS & $$-$0.859$ & $4.191$ & $$-$1.010$ & $$-$0.101$ & $0.348$ & $0.474$ & $0.355$ & $0.504$ \\ 
integrate G-estimation & $$-$1.320$ & $3.956$ & $$-$2.154$ & $$-$1.950$ & $0.348$ & $0.474$ & $0.360$ & $0.517$ \\ 
CE dWOLS & $$-$0.859$ & $4.191$ & $$-$1.010$ & $$-$0.101$ & $0.348$ & $0.474$ & $0.355$ & $0.504$ \\ 
CE G-estimation & $$-$1.320$ & $3.956$ & $$-$2.154$ & $$-$1.950$ & $0.348$ & $0.474$ & $0.360$ & $0.517$ \\ 
\hline \\[-1.8ex] 
dWOLS & $15.826$ & $10.723$ & $6.225$ & $12.738$ & $0.103$ & $0.141$ & $0.107$ & $0.150$ \\ 
G-estimation  & $15.815$ & $10.706$ & $6.212$ & $12.695$ & $0.103$ & $0.141$ & $0.107$ & $0.150$ \\ 
IPTW+dWOLS & $$-$0.557$ & $0.459$ & $0.121$ & $$-$0.063$ & $0.105$ & $0.139$ & $0.108$ & $0.149$ \\ 
IPTW+G-estimation & $$-$0.575$ & $0.454$ & $0.059$ & $$-$0.197$ & $0.106$ & $0.139$ & $0.108$ & $0.149$ \\ 
integrate dWOLS & $$-$0.527$ & $0.453$ & $0.086$ & $$-$0.173$ & $0.104$ & $0.139$ & $0.107$ & $0.149$ \\ 
integrate G-estimation  & $$-$0.546$ & $0.449$ & $0.064$ & $$-$0.225$ & $0.104$ & $0.139$ & $0.106$ & $0.149$ \\ 
CE dWOLS & $$-$0.527$ & $0.453$ & $0.086$ & $$-$0.173$ & $0.104$ & $0.139$ & $0.107$ & $0.149$ \\ 
CE G-estimation & $$-$0.546$ & $0.449$ & $0.064$ & $$-$0.225$ & $0.104$ & $0.139$ & $0.106$ & $0.149$ \\ 
\hline \\[-1.8ex] 
dWOLS & $16.414$ & $10.207$ & $6.276$ & $12.992$ & $0.034$ & $0.046$ & $0.033$ & $0.046$ \\ 
G-estimation & $16.413$ & $10.205$ & $6.270$ & $12.983$ & $0.034$ & $0.046$ & $0.033$ & $0.046$ \\ 
IPTW+dWOLS & $$-$0.130$ & $$-$0.063$ & $0.058$ & $0.017$ & $0.034$ & $0.045$ & $0.033$ & $0.045$ \\ 
IPTW+G-estimation & $$-$0.131$ & $$-$0.063$ & $0.049$ & $0.0001$ & $0.034$ & $0.045$ & $0.033$ & $0.045$ \\ 
integrate dWOLS & $$-$0.127$ & $$-$0.066$ & $0.049$ & $0.009$ & $0.033$ & $0.045$ & $0.033$ & $0.045$ \\ 
integrate G-estimation & $$-$0.128$ & $$-$0.067$ & $0.042$ & $$-$0.001$ & $0.033$ & $0.045$ & $0.033$ & $0.045$ \\ 
CE dWOLS & $$-$0.127$ & $$-$0.066$ & $0.049$ & $0.009$ & $0.033$ & $0.045$ & $0.033$ & $0.045$ \\ 
CE G-estimation & $$-$0.128$ & $$-$0.067$ & $0.042$ & $$-$0.001$ & $0.033$ & $0.045$ & $0.033$ & $0.045$ \\ 
\hline \\[-1.8ex] 
\end{tabular} }
\end{table} 

\begin{table}[!htbp] \centering 
  \caption{Proportion of the observations for which the optimal partially adaptive treatment strategy is correctly identified across replications} 
		\resizebox{\textwidth}{!}{	
\begin{tabular}{@{\extracolsep{5pt}} lccccccccc} 
\\[-1.8ex]\hline 
\hline \\[-1.8ex] 
& \multicolumn{3}{c}{Scenario 1} & \multicolumn{3}{c}{Scenario 2} & \multicolumn{3}{c}{Scenario 3} \\
n & 100 & 1000 & 10~000 & 100 & 1000 & 10~000 & 100 & 1000 & 10~000 \\ 
\hline \\[-1.8ex] 
dWOLS & $56.2$ & $50.9$ & $38.9$ & $55.2$ & $49.5$ & $49.8$ & $56.5$ & $51.2$ & $39.1$ \\ 
G-est & $56.6$ & $50.8$ & $38.9$ & $55.7$ & $49.2$ & $49.8$ & $56.8$ & $51.1$ & $39.1$ \\ 
IPTW+dWOLS & $60.4$ & $67.0$ & $80.5$ & $60.2$ & $65.8$ & $64.5$ & $61.0$ & $67.4$ & $79.5$ \\ 
IPTW+G-est & $61.1$ & $67.2$ & $80.4$ & $61.0$ & $65.5$ & $61.3$ & $61.3$ & $67.4$ & $79.5$ \\ 
integrate dWOLS & $61.1$ & $67.3$ & $80.2$ & $61.4$ & $67.7$ & $79.7$ & $61.2$ & $67.2$ & $79.4$ \\ 
integrate G-est & $61.6$ & $67.5$ & $80.4$ & $62.2$ & $68.5$ & $82.4$ & $62.2$ & $67.4$ & $79.5$ \\ 
CE dWOLS & $61.1$ & $67.3$ & $80.2$ & $61.4$ & $67.7$ & $79.7$ & $61.2$ & $67.2$ & $79.4$ \\ 
CE G-est & $61.6$ & $67.5$ & $80.4$ & $62.2$ & $68.5$ & $82.4$ & $62.2$ & $67.4$ & $79.5$ \\ 
\hline \\[-1.8ex] 
\end{tabular} }
\end{table} 

\clearpage

\section*{Appendix F - Additional simulation to investigate confidence intervals}

\subsection*{Scenario}
We performed additional simulations to explore the ability of the $m$-out-of-$n$ bootstrap to yield adequate confidence intervals for the parameters of a PATS. Because of the important computational burden of this type of bootstrap, we considered a single simulation scenario with two time-points, Scenario 1 of Appendix E, and a single estimator, ``CE dWOLS.'' We generated 1000 datasets of size $n = 300$. 

\subsection*{Confidence intervals}
Confidence intervals were constructed using the $m$-out-of-$n$ where $m$ was chosen data-adaptively using the procedure proposed by Chakraborty et al (2013). Briefly, a first standard non-parameteric ($n$-out-of-$n$) bootstrap was first performed with $B_1 = 500$ replicates. Using these, we estimated the variance-covariance matrix of the estimators of the parameters of the blip at the second time-point ($\hat{\psi}_2^*$). We then constructed 95\% confidence intervals for the linear predictor $a_2 h_2 \hat{\psi}_2^*$ for all observations. We estimated a non-regularity parameter $p$ as the proportion of the observations for which the optimal treatment at the second time-point was not uniquely defined, that is, the proportion of the observations for which the preceding confidence intervals included the null value. 

Next, we took $B_2 = 500$ samples with replacement of size $m = n^{\frac{1+\alpha(1 - \hat{p})}{1 + \alpha}}$ within each first-stage bootstrap sample, where $\alpha = 0.025$ and $\hat{p}$ is the estimated value of $p$. For each of the $B_1$ first stage replicate, 95\% confidence intervals for the parameters of the blip were computed as the $2.5^{th}$ and $97.5^{th}$ percentiles of the $B_2 = 500$ second-stage bootstrap replicates. As such, $B_1 = 500$ confidence intervals were computed for each parameter. We computed the proportion of these confidence intervals that included their respective first-stage bootstrap estimate. If this proportion was greater than $95\%$ then $\hat{\alpha} = \alpha$ and $\hat{m} = m$. Otherwise we increased the value of $\alpha$ by 0.025 and restarted the second-stage bootstrap with the new corresponding value of $m$. This procedure was repeated until at least 95\% of the confidence intervals produced at the second stage included the first stage estimates. Note that if $\hat{p} = 0$, then $\hat{m} = n$ and no second-stage bootstrap was performed. Once the procedure stopped, we performed one last non-parametric bootstrap with $B_1 = 500$ replicates of size $\hat{m}$ and 95\% confidence intervals were based on the $2.5^{th}$ and $97.5^{th}$ percentiles of the replicates. 

\subsection*{Results}

The value of $\hat{p}$ varied between 0 and 1, with a mean of 0.47. When $\hat{p} \neq 0$, $\hat{\alpha}$ was either 0.025 or 0.05 in 81.5\% of the simulation replicates. Its mean value was 0.04 and the maximum was 0.125. When $\hat{p} \neq 0$, each $\alpha$ value explored took about 50 minutes of computation time. The sample size of the $m$-out-of-$n$ bootstrap varied between 217 and 300, with a mean of 269. The coverage for all four PATS parameters was approximately 95\%: 95.7\% for $\psi^*_{10}$, 95.2\% for $\psi^*_{11}$, 95.2\% for $\psi^*_{20}$ and 95.1\% for $\psi^*_{21}$.  

\end{document}